# Phytoplankton temporal strategies increase entropy production in a marine food web model

**Joseph J. Vallino** [1] * and **Ioannis Tsakalakis** [2]

[1] Marine Biological Laboratory, Woods Hole; jvallino@mbl.edu
[2] Marine Biological Laboratory, Woods Hole and Massachusetts Institute of Technology, Cambridge; itsakalakis@mbl.edu
* Correspondence: jvallino@mbl.edu

**Abstract:** We develop a trait-based model founded on the hypothesis that biological systems evolve and organize to maximize entropy production by dissipating chemical and electromagnetic potentials over longer time scales than abiotic processes by implementing temporal strategies. A marine food web consisting of phytoplankton, bacteria and consumer functional groups is used to explore how temporal strategies, or the lack there of, change entropy production in a shallow pond that receives a continuous flow of reduced organic carbon plus inorganic nitrogen and illumination from solar radiation with diel and seasonal dynamics. Results show that a temporal strategy that employs an explicit circadian clock produces more entropy than a passive strategy that uses internal carbon storage or a balanced growth strategy that requires phytoplankton to grow with fixed stoichiometry. When the community is forced to operate at high specific growth rates near 2 d$^{-1}$, the optimization-guided model selects for phytoplankton ecotypes that exhibit complementary for winter versus summer environmental conditions to increase entropy production. We also present a new type of trait-based modeling where trait values are determined by maximizing entropy production rather than by random selection.

## 1. Introduction

The maximum entropy production (MEP) principle postulates that steady state, non-equilibrium systems with many degrees of freedom will likely organize to maximize entropy production, or equivalently, maximize the dissipation rate of energy potentials [1-3]. MEP applications can be traced back to at least Paltridge [4], and perhaps to even Lotka [5], and MEP theory appears to have multiple origins [4,6-8], but over the last decade and a half there has been increasing interest in extending MEP theory as well as its applications [9,10]. Since MEP makes no distinction between abiotic or biotic systems, MEP research has been wide ranging, from crystal growth [11], Rayleigh-Benard convection [12], phase transitions [13] to Earth's hydrologic cycle [14], ocean circulation [15], ecology [16], biogeochemistry [17] and evolution [18] to name just a few. The MEP approach has garnered interest in systems where classical reductionist modeling is difficult to implement due insufficient information or understanding, such as turbulent flow and living systems that are governed by self-organization. For systems where we have good reductionist understanding and modeling capabilities (i.e., Newtonian physics), MEP may provide some benefit from a wholistic perspective, but otherwise may not be of much use.

One of the important uncertainties regarding MEP theory concerns its applicability to non-steady state systems for which no theory currently exists. There have been some attempts to extend MEP theory to non-steady state systems [19,20], but no consensus has been attained. Consequently, we have chosen a different approach to exploring MEP in non-steady state systems based on numerical modeling of the chemistry catalyzed by microbial systems and comparing modeling results to observations or basic understanding of these systems. If MEP-based models show good predictive capabilities, then the likelihood that MEP governs their function and evolution is supported. This is particularly relevant for biological systems as we have yet to develop good predictive understanding of their behavior due to their complexity and self-organizing capabilities. Bacteria and archaea have evolved the capability to exploit redox chemical potentials found in numerous environments on Earth, such as oxidation of hydrogen sulfide or ammonium with oxygen while fixing carbon dioxide (chemolithoautotrophy), oxidation of reduced organic carbon with sulfate or nitrate (anerobic metabolism), and oxidation of hydrogen with carbon dioxide to synthesize methane (methanogenesis) to name just a few [21]. All of these exergonic reactions occur abiotically, but the presence of biology can increase reaction rates and associated destruction of chemical potential by many orders of magnitude. Modeling this biogeochemistry using reductionism is challenging because a litter of water contains $10^9$ or more individuals, thousand to perhaps 10 of thousands of different species [22], all of which are subject to predation and viral attack; soils and sediments are approximately 1000 times more complex. While it may not be possible to model all the details of these communities and the associated chemistry they catalyze, MEP provides an opportunity for prediction, assuming living systems evolve, organize and function to dissipate energy potentials. If MEP theory does not explain microbial systems, there seems little expectation that it would be useful in describing biogeochemistry of higher trophic levels where the theory of large numbers is even less applicable [23], so microbial systems are a good place to focus effort.

Our approach to modeling microbial biogeochemistry uses a distributed metabolic network [24] to represent the possible reactions a microbial community as a whole can catalyze, with emphasis placed on compounds found in the ecosystem rather than those found within a cell. The reactions included in the network are divided into those that release Gibbs free energy (exergonic or catabolic reactions) and those associated with biomass synthesis (endergonic or anabolic reactions), where biomass is considered more as a catalyst than an organism. In fact, the catalyst associated with a given redox reaction can represent numerous species capable of conducting the reaction but are not distinguished in the model. Initial research on this approach focusing on non-steady state systems revealed that the time scale over which entropy is maximized changes the solution significantly [25]. In particular, if entropy production (EP) is maximized instantaneously then no biomass is synthesized, but the solution does exhibit characteristics of abiotic processes, while if EP is maximized over a given time interval, then the solutions are consistent with the actions of biology, and more entropy can be produced. Based on this initial work [17,26,27], our current working hypothesis is that living systems use information acquired by evolution and stored in the genome to maximize destruction of energy potentials over time scales where prediction fidelity is sufficient. This capability of biology can lead to greater EP over time, which differs from abiotic systems that maximize EP instantaneously, such as fire or a rock rolling down a hill. That is, abiotic systems take steepest descent routes down free energy surfaces. While not a focus of this manuscript, the working hypothesis can be extended to space as well, where abiotic systems maximize local EP and biological systems maximize EP integrated over space by coordinating function via communication, such as quorum sensing [28].

One of the hallmarks of biology is the evolution of temporal strategies, such as circadian rhythms [29,30], food and resource storage [31,32], anticipatory control [33,34] and dormancy [35], which are consistent with the MEP hypothesis regarding abiotic versus biotic systems; however, such temporal strategies are seldom incorporated into models describing biogeochemistry of microbial processes. The objectives of this manuscript are to 1) illustrate an approach for incorporating temporal strategies into biogeochemical models, 2) test a new approach that replaces our previously developed optimal control approach with trait-based modeling where trait values are determined by MEP optimization and 3) show that inclusion of temporal strategies leads to greater EP. We use a marine microbial food web as an example system that consists of photoautotrophs (aka, phytoplankton), aerobic heterotrophic bacteria that consume dissolve organic matter, and general consumers that can prey on both phototrophs and heterotrophs, as well as cannibalize themselves.

## 2. Model Description

The overall objective of the model is to explore how the addition of phytoplankton temporal strategies alter entropy production (EP) in marine planktonic food webs that dissipate electromagnetic and chemical potentials from solar radiation and reduced organic carbon inputs, such as glucose. We also investigate a new hybrid approach that uses EP maximization to set trait values in a trait-based modeling approach. The details and governing equations for the model are described in detail in *Supplementary Materials*, but key aspects of the model are described here. Double bracket notation, ⟦ ⟧, will be used to denote a variable's units in the model.

*2.1 Entropy production*

As discussed elsewhere [5,36,37], EP in this manuscript refers solely to irreversible processes and does not refer to system or state entropy, which traditionally is represented by the symbol $S$. We follow convention here and use $\dot{\sigma}$ ⟦$J\ d^{-1}\ K^{-1}$⟧ to represent EP from irreversible processes, and $\sigma$ ⟦$J\ K^{-1}$⟧ to represent cumulative entropy produced from irreversible processes over a specified time interval, so that $\sigma = \int_{t}^{t+\Delta t} \dot{\sigma}\, d\tau$. Thermodynamic entropy production is the destruction of energy potentials or Gibbs free energy; contrary to popular believe, system order contributes little to $\dot{\sigma}$ for biological systems or the chemistry they catalyze [38]. For chemical and biological systems, entropy production is readily calculated from the product of the reaction rate, $r$ ⟦$mmol\ m^{-3}\ d^{-1}$⟧, times the Gibbs free energy of reaction, $\Delta_r G$ ⟦$J\ mmol^{-1}$⟧, divided by temperature, as given by,

$$\dot{\sigma} = -\frac{1}{T} r \Delta_r G \qquad (1)$$

For non-steady state systems, maximizing $\sigma$ depends on the time interval, $\Delta t$. For instance, in biological systems if entropy production is maximized instantaneously, then no organismal growth can occur. Consider a system that starts with some bacterial biomass present, along with some chemical potential, such as glucose plus oxygen. The maximum instantaneous EP is attained by maximizing the rate of glucose oxidation, as any co-synthesis of biomass would reduce EP. Furthermore, if a bacterial consumer is present, then instantaneous EP can be increased further by oxidizing the bacteria as well. Hence, under instantaneous EP, all chemical potentials get destroyed, which is analogous to fire. However, if EP is maximized over some time interval, such a day, then synthesizing biomass can result in greater EP over the interval because reaction rate is proportional to biomass concentration, and entropy production is proportional to reaction rate, as given by Eqn. (1). Mathematically, this can be expressed as,

$$\max\left(\int_{t}^{t+\Delta t} \dot{\sigma}(\tau)\, d\tau\right) \geq \int_{t}^{t+\Delta t} \max\left(\dot{\sigma}(\tau)\right) d\tau, \qquad (2)$$

where the left-hand side of Eqn. (2) represent biology and the temporal strategies it implements, while the right-hand side of Eqn. (2) represents instantaneous, or steepest decent trajectory, abiotic systems follow, but see [5,25-27] for further discussion. Eqn. (2) is the fundamental hypothesis of the model.

*2.2 Metabolic reactions*

The model, derived in part from previous work focused on the biogeochemistry in a meromictic pond [17], consists of three functional groups: phytoplankton, $\mathcal{S}_P$ ⟦$mmol\ m^{-3}$⟧, which intercept high frequency photons, $\gamma_H$ ⟦$mmol\text{-}\gamma$⟧, and fix $CO_2$ producing $O_2$; bacteria, $\mathcal{S}_B$, that consume labile organic carbon, $C_L$, and decompose recalcitrant organic carbon, $C_R$, and nitrogen, $N_R$, into labile constituents; consumers, $\mathcal{S}_C$, that prey on phytoplankton and bacteria, as well as themselves and produce recalcitrant organic carbon and nitrogen (Figure 1). Unlike our previous models that contained just one state variable for each functional group, in the trait-based approach there are $n_P$, $n_B$ and $n_C$ instances or ecotypes of $Ṣ_{P\{i\}}$, $Ṣ_{B\{i\}}$, and $Ṣ_{C\{i\}}$, respectively, where a particular ecotype is distinguished using braces nomenclature, such as $P\{i\}$. The symbol $Ṣ$ represents biological $Ṣ$tructure to emphasize its action as a reaction catalyst as opposed to the organismal centric view typically pursued in biology. The generalized reactions each functional group catalyzes are listed in Table 1, and the stoichiometrically balanced reactions are provided in Section S2.2 of the *Supplementary Material*. Two different kinetics expressions govern growth of phototrophs, $Ṣ_P$, one of which is also used for heterotrophs, $Ṣ_B$, $Ṣ_C$, as described Section 2.3 below.

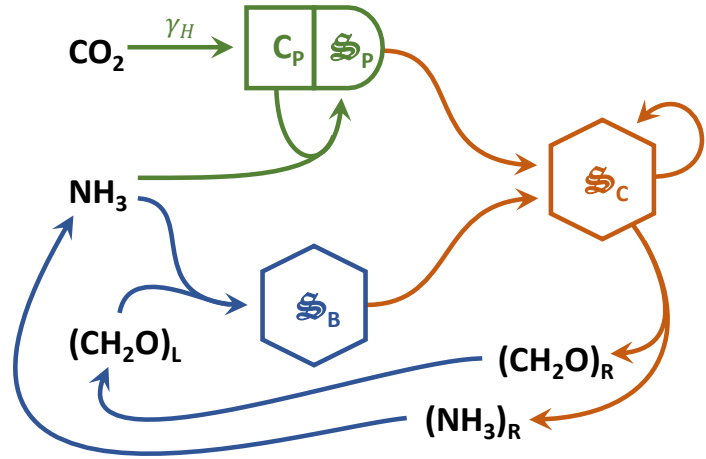

Figure 1. Food web structure used in the marine plankton model consisting of three function groups representing phytoplankton, bacteria and consumers. Colored lines correspond to reactions a functional group is capable of catalyzing.

Table 1. Reactions associated with the three functional groups (phytoplankton ($P$), bacteria ($B$) and consumers ($C$)), where $r_{i,\chi\{j\}}$ represents sub-reaction $i$ of biological catalyst $Ṣ_{\chi\{j\}}$, and $\chi\{j\}$ is ecotype $\{j\}$ of $P$, $B$ or $C$. For consumers, $i$ spans all $P\{i\}$, $B\{i\}$ and $C\{i\}$; consequently, among the 3 functional groups there are a total of $2n_P + 3n_B + (n_P + n_B + n_C)n_C$ reactions. Reactions are shown here to emphasize function only. Complete reaction stoichiometries are provided in Section S2.2 of the *Supplementary Material* and $H_3PO_4$ and $P_D$ are only used in thermodynamic calculations but are not state variables in the model.

| Rxn. | Abbreviated Stoichiometry | Cat. |
|---|---|---|
| $r_{1,P\{j\}}$ | $H_2CO_3 + \gamma_H \rightarrow C_{P\{j\}} + O_2$ | $Ṣ_{P\{j\}}$ |
| $r_{2,P\{j\}}$ | $C_{P\{j\}} + NH_3 + H_3PO_4 + O_2 \rightarrow Ṣ_{P\{j\}} + H_2CO_3$ | $Ṣ_{P\{j\}}$ |
| $r_{1,B\{j\}}$ | $C_L + NH_3 + H_3PO_4 + O_2 \rightarrow Ṣ_{B\{j\}} + H_2CO_3$ | $Ṣ_{B\{j\}}$ |
| $r_{2,B\{j\}}$ | $C_D \rightarrow C_L$ | $Ṣ_{B\{j\}}$ |
| $r_{3,B\{j\}}$ | $N_D \rightarrow NH_3$ | $Ṣ_{B\{j\}}$ |
| $r_{\chi\{i\},C\{j\}}$ | $Ṣ_{\chi\{i\}} + C_{\chi\{i\}} + O_2 \rightarrow Ṣ_{C\{j\}} + H_2CO_3 + C_D + NH_3 + N_D + H_3PO_4 + P_D$ | $Ṣ_{C\{j\}}$ |

To remove parameters whose values are largely unknown and poorly bounded, we formulate growth kinetics to depend on just two types of control variables whose values are determined over space and time to maximize EP, so that there are no parameters that require tuning. Consequently, the MEP approach requires very little information other than the biogeochemical reactions the community is capable of catalyzing and the constraints the environment places on the system, which is largely in the form of transport processes that govern free energy and resources input and output to the system. A metabolic reaction for a microorganism is constructed from a combination of two reaction types, one associated with the extraction of Gibbs free energy from chemical or electromagnetic potentials in the environment, and the other associated with biosynthesis of biological structure, $\mathfrak{S}$ (or catalyst), driven by the extracted free energy. This representation derives from biochemistry where catabolic reactions produce ATP to drive anabolic reactions. The two reactions are combined to give a single reaction in the metabolic network. For phototrophs, two reactions are used, which are approximately given by,

$$r_{1,P}: \varepsilon_P H_2CO_3 + n_{1,P}\gamma_H \xrightarrow{\Omega_{1,P}\mathfrak{S}_P} \varepsilon_P(C_P + O_2) \tag{3}$$

and

$$r_{2,P}: C_P + \varepsilon_P \gamma NH_3 + a(1-\varepsilon_P)O_2 \xrightarrow{\Omega_{2,P}\mathfrak{S}_P} \varepsilon_P \mathfrak{S}_P + bH_2O + (1-\varepsilon_P)CO_2. \tag{4}$$

In the first reaction, Gibbs free energy of high frequency photons, $\gamma_H$, from photosynthetic active radiation (PAR) is used to drive the anabolic reaction of converting $CO_2$ plus water into reduced carbon, $C_P$ (glucose on a unit carbon basis) and oxygen (Eqn. (3)). In the second reaction, the reduced carbon is combined with elemental resources (only ammonium shown here) to produce phytoplankton biomass, $\mathfrak{S}_P$, where some of the reduced carbon is oxidized to $CO_2$ and $H_2O$ to drive biosynthesis (Eqn. (4)). Both reactions are catalyzed by $\mathfrak{S}_P$, so neither proceed in its absence. The two control variable classes in these reactions, as well as all reactions in the network (Figure 1, Table 1), are a growth or thermodynamic efficiency variable, $\varepsilon_P$, and a set of cellular resource allocation variables, $\Omega_{1,P}$ and $\Omega_{2,P}$. The efficiency variable determines how near thermodynamic reversibility a reaction proceeds. When $\varepsilon_P = 1$, all of the electromagnetic and chemical potentials are conserved and just converted to other chemical potentials ($C_P + O_2$ in the first reaction and $\mathfrak{S}_P$ in the second). In theory, these reactions could be reversed to reproduce the constituents on the left-hand-side of the reactions. In this case, no entropy is produced, but reaction rates are zero due to the thermodynamic reversibility requirement or lack of thermodynamic force [39,40]. On the contrary, when $\varepsilon_P = 0$, both potentials are completely destroyed, and maximum entropy production occurs, but no $C_P$ nor catalyst are produced. The control variables $\Omega_{1,P}$ and $\Omega_{2,P}$ determine how much of the catalyst, $\mathfrak{S}_P$, is allocated to each reaction. Since catalyst must be conserved, one degree of freedom can be removed because $\Omega_{1,P} + \Omega_{2,P} = 1$. Consequently, to maximize entropy production overall, the optimization objective is to synthesize just enough catalyst given recourse available to maximize the dissipation of the energy potentials, which is the essence of the MEP theoretical and modeling approach.

For chemotrophic bacteria and heterotrophic consumers, reactions are similar to that given by Eqn. (4), except more reactions can be added to the functional groups. For example, in addition to the equation for bacterial growth, $r_{1,B\{i\}}$ (Table 1), bacteria also have the capability to decompose recalcitrant detrital carbon, $C_D$, and nitrogen, $N_D$, into labile pools via $r_{2,B\{i\}}$ and $r_{3,B\{i\}}$, respectively (Table 1). The amount of catalyst allocated to the three reactions is governed by the control variables $\Omega_{1,B\{i\}}$, $\Omega_{2,B\{i\}}$ and $\Omega_{3,B\{i\}}$, respectively, but these control variables must sum to 1, so only two degrees of freedom are needed to determine the partitioning of $\mathfrak{S}_{B\{i\}}$ to the three reactions. Gibbs free energy of reaction for the two decomposition reactions is calculated from the logarithm of the concentration differences between reactants and products, as the Gibbs free energy of formation is assumed to be equal for labile and recalcitrant C and N (see Eqns. (S41 and S46) in Section S2.2.2).

For the heterotrophic consumers, which can consume all prey including other consumers and themselves, each consumer has $n_P + n_B + n_C$ prey consuming reactions, where $\Omega_{\chi\{j\},C\{i\}}$ determines allocation of catalyst, $\mathfrak{S}_{C\{i\}}$, to each prey, and $\chi\{j\}$ is either $P\{j\}$, $B\{j\}$ or $C\{j\}$. Unlike phytoplankton and bacteria, it is assumed consumers allocate catalytic resources to prey based on the preys concentration relative to all prey, so reaction rates depend on a weighted version of $\Omega_{\chi\{j\},C\{i\}}$ denoted as $\omega_{\chi\{j\},C\{i\}}$ as defined by Eqn. (S56). The predation by consumers also produces labile nitrogen as well as recalcitrant carbon on nitrogen as a function of $\varepsilon_{C\{i\}}$, Eqn. (S50), to capture "sloppy feeding" and incomplete digestion [41]. Furthermore, when consumers prey on phytoplankton, it is assumed that the labile carbon they contain, $C_{P\{i\}}$, is combusted and not incorporated into catalyst but does contribute to Gibbs free energy of reaction, as given by Eqns. (S31-S35). Reaction stoichiometry of consumers has based on several assumptions along with the principal objective of not introducing any new parameters other than the control variables $\varepsilon_{C\{i\}}$ and $\Omega_{\chi\{j\},C\{i\}}$.

*2.3 Reaction kinetics*

Two types of kinetic expressions, one for electromagnetic potentials and another for chemical potentials, govern reaction rates in the model, which were developed in Vallino and Huber [17], but are briefly describe here. Phototrophs incorporate both types of kinetics, as the first reaction (Eqn. (3)) utilizes electromagnetic radiation to drive $C_P$ synthesis, while the chemical oxidation of $C_P$ is used for $S_P$ synthesis in Eqn (4). Photosynthetic radiation enters a system on an areal basis, which differs from chemical potentials that are volume based. Chemical reaction rates depend on the state of the system—reactant concentrations—, while photoreaction rates depend on the rate of areal photon input; consequently, photosynthetic reaction rates are proportional to $I(t, z)$ $[\![mmol\text{-}\gamma\ m^{-2}d^{-1}]\!]$, which is the light intensity at time $t$ and water depth $z$.

The rate at which high frequency photons, $\gamma_H$, are captured by phytoplankton photosynthetic machinery per unit volume, $\Delta I_{P\{i\}}$ $[\![mmol\text{-}\gamma\ m^{-3}d^{-1}]\!]$, is given by,

$$\Delta I_{P\{i\}} = k_{Chl}\Omega_{1,P\{i\}}[S_{P\{i\}}]\langle I(t)\rangle_d \tag{5}$$

where $k_{Chl}$ $[\![m^2\ (mmol\text{-}C)^{-1}]\!]$ is the light attenuation coefficient by chlorophyll a [42], $\Omega_{1,P\{i\}}[S_{P\{i\}}]$ is the fraction of phytoplankton biomass allocated to photosynthesis and $\langle I(t)\rangle_d$ is the depth-averaged light intensity $[\![mmol\text{-}\gamma\ m^{-2}d^{-1}]\!]$ for a well-mixed column of water of depth $d$ $[\![m]\!]$ (Eqn (S9)). Note for simplicity, we only consider blue light of 440 nm wavelength, as a full spectrum model would require considerably more development and is beyond the scope of this study.

In Eqn. (3), $n_{1,P}$ is the mmoles of photons needed to fix one mmole of CO$_2$ under thermodynamic reversibility (see Eqn. (S6)), so the maximum rate of reaction $r_{1,P}$ is given by $\frac{\Delta I_{P\{i\}}}{n_{1,P\{i\}}}$. However, the reaction rate can also be limited by CO$_2$ plus HCO$_3^-$ concentration, so the overall rate expression for photon driven $r_{1,P\{i\}}$ reaction is given by,

$$r_{1,P\{i\}} = \frac{\Delta I_{P\{i\}}}{n_{1,P\{i\}}}\left(\frac{[CO_2] + [HCO_3^-]}{[CO_2] + [HCO_3^-] + \kappa^*\varepsilon_{P\{i\}}^4}\right)F_T(\Delta_r G_{1,P\{i\}}, n_{1,P\{i\}}^e) \tag{6}$$

where $F_T$ is a thermodynamic force as described by LaRowe et al. [40], which depends on the Gibbs free energy of reaction and the number of electrons transferred, $n_{1,P\{i\}}^e$ (also see [17]). Simply stated, as the reaction free energy goes to zero, $F_T$ drives the reaction rate to zero regardless of how favorable the kinetics are. The form of the kinetic force, in particular the $\kappa^*\varepsilon_{P\{i\}}^4$ term in the Monod-like expression, will be explained next, as it was developed for reaction kinetics driven by chemical potentials [28].

The second class of reaction kinetics used for chemotrophs (i.e., bacteria) and consumers, as well as for the phytoplankton biosynthesis reaction, $r_{2,P\{i\}}$, is an adaptive Monod equation, which has the general form given by,

$$r_{j,\chi\{i\}} = \nu^*\varepsilon_{\chi\{i\}}^2\Omega_{j,\chi\{i\}}[S_{\chi\{i\}}]F_K(\mathbf{c},\mathbf{\Lambda}_{j,\chi},\varepsilon_{\chi\{i\}})F_T(\Delta_r G_{j,\chi\{i\}}, n_{j,\chi\{i\}}^e), \tag{7}$$

where the kinetic force, $F_K(\mathbf{c},\mathbf{\Lambda}_{j,\chi},\varepsilon_{\chi\{i\}})$, is given by,

$$F_K(\mathbf{c},\mathbf{\Lambda}_{j,\chi},\varepsilon_{\chi\{i\}}) = \prod_{k=1}^{n_S}\left(\frac{c_k}{c_k + \kappa^*\varepsilon_{\chi\{i\}}^4}\right)^{\Lambda_{j,k,\chi}}. \tag{8}$$

The maximum biomass-specific reaction rate, $\nu^* \varepsilon_{\chi\{i\}}^2 \; [\![d^{-1}]\!]$, in Eqn. (7) and the half saturation constant, $\kappa^* \varepsilon_{\chi\{i\}}^4 \; [\![mmol \; m^{-3}]\!]$, in the kinetic force, Eqn. (8), are both parameterized by $\varepsilon_{\chi\{i\}}$. The fixed parameters $\nu^*$ and $\kappa^*$ have been chosen so that as $\varepsilon_{\chi\{i\}}$ varies from 0 to 1, the growth kinetics describe a family of curves that represent the growth of oligotrophs to copiotrophs, respectively [28]. Since $\nu^*$ and $\kappa^*$ are held constant at 350 d$^{-1}$ and 5000 mmol m$^{-3}$, respectively, for all functional groups and reactions, except detritus decomposition, there are no adjustable parameters other than the two control variables $\varepsilon_{\chi\{i\}}$ and $\Omega_{j,\chi\{i\}}$ governing reaction rates and stoichiometry. For decomposition of recalcitrant organic matter given by $r_{2,B\{i\}}$ and $r_{3,B\{i\}}$, $\nu_D^*$ replaces $\nu^*$ to reflect the slower kinetics associated with detritus utilization, where $\nu_D^*$ is set to 175 d$^{-1}$. The kinetic force, $F_K$, depends on the concentrations, $c_k \; [\![mmol \; m^{-3}]\!]$, of the state variables of which there are $n_S$, and $\Lambda_{j,\chi}$ is a binary matrix that specifies the reactants used in reaction $j$ of biological structure $\chi$, which can be $P, B$ or $C$, where $\Lambda_{j,k,\chi}$ equals 1 if reactant $c_k$ is used in reaction $r_{j,\chi}$; otherwise, $\Lambda_{j,k,\chi}$ equals 0.

Equations (7) and (8) also govern consumer, $\mathfrak{S}_{C\{i\}}$, reaction rates (see Eqn. (S57)), but the number of "reactants" is the total number of prey in the simulation ($n_P + n_B + n_C$), so that the resource allocation variable, $\Omega_{\chi\{j\},C\{i\}}$, is a matrix of traits with size $\mathbb{R}^{(n_P+n_B+n_C)\times n_C}$. This matrix grows rapidly as the number of ecotypes in the model are increased, which introduces some challenges in exploring trait space as discussion in Section 2.5 below. The details for each reaction are provided in the *Supplementary Material*.

*2.4 Model domain and simulation details*

For simplicity, the model domain is a pond-like cylindrical reservoir of depth $d$ with unit surface area that is illuminated at the surface with both diel and seasonal solar radiation and can exchange O$_2$ and CO$_2$ with the atmosphere. The pond is modeled as a well-mixed chemostat (0D) with equal input and output flows, $F$, and a fixed volume, $V$, which defines a dilution rate given by $D = \frac{F}{V}$. A governing set of ordinary differential equations (ODEs, Eqn. S1) is derived from mass balances (Section S2.3) around the 6 constituents (DOC, O$_2$, C$_L$, C$_D$, N$_D$ and NH$_3$) and the $n_P$ phytoplankton, $n_P$ phytoplankton carbon stores, $n_B$ bacteria and $n_C$ consumers. An additional three ODEs (Eqns S70-S72) integrate irreversible entropy production, $\dot{\sigma}$, to obtain total entropy production, $\sigma^T$, with contributions from reactions, $\sigma^R$, particles, $\sigma^P$, and water, $\sigma^W$. All simulations are run for two years with constant inputs at a dilution rate of 0.2 d$^{-1}$ unless otherwise specified.

*2.5 Optimize Trait-based model*

In previous work, we have used optimal control to determine how the growth kinetics of each functional group changes over both time and space to maximize $\sigma$ [17,25,26,28]. In that approach only a single state variable is used to represent each functional group, but the growth characteristics of each group can change over space and time as dictated by the control variables. A disadvantage of the approach is that the dimension of the optimal control problem grows rapidly with each spatial dimension added and becomes computationally prohibitive for 2D and 3D spaces, at least under the current numerical approach. Even though the model developed here is 0D, we explore a new approach that uses trait-based modeling which can be extended to 2D and 3D environments.

In trait-based modeling [43], parameters are considered as traits, reflecting optimal growth conditions, such as temperature and light intensity, that incorporate appropriate tradeoffs, such as high specific growth rate but low substrate affinity. The model domain is populated with many ecotypes in each functional group and trait values are randomly selected from an appropriate distribution, so that models start with high biodiversity. The models can have hundreds of state variables capturing diverse ecotypes, but as model simulations proceed *in silico* natural selection culls ecotypes from the population that have poor growth kinetics for the local environmental conditions. Furthermore, by continuously seeding the simulation with low concentrations of all ecotypes, if the simulated environment changes, new ecotypes can be selected for, which removes some of the problems that plague classic biochemistry models regarding population restructuring and the need to recalibrate parameters. To achieve good numerical coverage of a particular trait, such as substrate affinity, a reasonable number of ecotypes with that trait must be included in the simulation, so a large number of traits can lead to a computationally prohibitive number of state variables, which is particularly true when consumers are included in the network.

As mentioned above, the consumer trait matrix $\Omega_{\chi\{j\},C\{i\}} \in \mathbb{R}^{(n_P+n_B+n_C)\times n_C}$ determines prey preferences for consumer $\mathbf{\mathfrak{s}}_{C\{i\}}$, but it also determines the trophic structure of the food web, because consumer $C\{i\}$ can consume consumer $C\{j\}$, which in turn could consume another consumer to produce a four-level trophic food web. More importantly, as the number of phytoplankton, bacteria and consumer ecotypes are added to the model to explore trait space, the matrix size increases roughly as the square of the number of prey. For instance, if 10 ecotypes are used for $P$ and $B$, then a single consumer will have 21 traits (row of $\Omega_{\chi\{j\},C\{i\}}$), which would need to be explored by adding more consumers with different values for the 21 traits, but adding more consumers increase the column dimension of $\Omega_{\chi\{j\},C\{i\}}$ and the number of traits used by each consumer. Consequently, in this configuration the traits space for $\mathbf{\mathfrak{s}}_{C\{i\}}$ increases faster than it can be explored. One solution is to limit the number of prey a consumer can have, but this places strong constraints on the structure of the food web, which is something we wanted to avoid in this study. This conundrum lead to the development of a new optimal trait-based modeling approach.

Instead of exploring trait space by adding many ecotypes to a model and relying on *in silico* natural selection to find the best trait values, we used a hybrid approach between trait-based modeling and our previous optimal control modeling. In this new approach, we populate the model with relatively few ecotypes for each functional group, then use optimization to search for trait values that maximize entropy production over a specified time interval instead of randomly assigning values.

All simulations were run with a fixed depth of 1 m, unless otherwise noted, because a deeper water column would require use of a 1D transport model so that local entropy production could be optimized at each depth. While dissolved constituents are assumed to be well mixed, this does not apply to light, which exhibits an exponential decrease in intensity as a function of depth and the concentration of particles and Chl a. As discussed in Vallino and Huber [17], entropy production associated with dissipation of electromagnetic potentials, or any energy potential that is quickly dissipated abiotically, must be maximized locally instead of globally to obtain simulations that are consistent with biology. This can be achieved by discretizing the water column into thin, ~ 1 m, depths, then conducting EP optimization in each layer; consequently, simulations presented here are just one layer at the surface.

*2.6 Temporal strategies*

Three phytoplankton temporal strategies are investigated here that affect how phytoplankton biomass is allocated to the two associated reactions, $r_{1,P\{i\}}$ and $r_{2,P\{i\}}$ (Table 1): 1) balanced growth; 2) passive C storage; 3) circadian resource allocation. For the first case with no temporal strategy, it is assumed that phytoplankton are limited to balanced growth, so that the ratio of $\mathbf{\mathfrak{s}}_{P\{i\}}$ to $C_{P\{i\}}$, or C:N ratio of phytoplankton, remains constant during the simulation, which is achieved by constraining $r_{1,P\{i\}}$ and $r_{2,P\{i\}}$ as given by Eqns. (S79-S81) and described in Section S3.2. For passive storage, $r_{1,P\{i\}}$ and $r_{2,P\{i\}}$ are not coupled, which allows $C_{P\{i\}}$ to increase and decrease based on mass balance in a manner reminiscent of Droop's formulation [44]. However, in passive storage there is no temporal variation in resource allocation, so $\Omega_{1,P\{i\}}$ and $\Omega_{2,P\{i\}}$, remain constant for the duration of the simulation. In circadian allocation, the resource allocation trait, $\Omega_{1,P\{i\}}$, can vary with time. Initial studies used a sinusoid functional for $\Omega_{1,P\{i\}}(t)$, Eqn. (S74), where frequency, $f_{P\{i\}}$, and phase, $\varphi_{P\{i\}}$, were used as traits and determined by EP maximization along with all other traits. While this approach worked, the global optimum was always found to be $f_{P\{i\}} = 1\,d^{-1}$, but it was computationally difficult to locate due to the narrowness of the optimum (Figure S1). To increase computational speed, we choose a square-wave function for $\Omega_{1,P\{i\}}(t)$, Eqn (S76), that varies on a diel cycle, where three trait parameters, $t_{On\{i\}}$, $t_{Off\{i\}}$ and $\Omega_{amp\{i\}}$ are used to set the time of step-up, step-down and amplitude of the square wave, respectively. Setting $t_{On\{i\}}$ and $t_{Off\{i\}}$ to 0 and 1 d$^{-1}$, respectively, produces the same results as the passive storage strategy.

*2.7 Optimization and computational approach*

As discussed in Section S3, all simulations were conducted on a small cluster (90 CPUs), where each CPU solved the optimization problem from an initial location in trait space that was selected from sampling a Latin hypercube to facilitate locating the global maximum. The basic algorithm is as follows. The optimization routine (hyperBOB) maximizes total entropy production, $\sigma^T$, by setting the values of the trait variables ($\varepsilon_{\chi\{i\}}$, $\Omega_{j,\chi\{i\}}$, $t_{On\{i\}}$, $t_{Off\{i\}}$ and $\Omega_{amp\{i\}}$) and passing them to the ODE integrator (BiM [45]) which determines how the state variables and entropy production vary over a two year period. Total entropy produced over the two-year period is returned to the optimization routine, that then adjusts the values of the trait variables to maximize entropy production based on a quadratic reconstruction of the optimum surface (see BOBYQA [46]). Iteration ends once the search region decreases to a user specified minimum radius.

Most of the simulations run in this study used a $1P1B1C$ food web model, but we also explored other small networks. In addition to model runs that explored the three temporal strategies described in Sections 2.6 and S3.2, we also examine how $\sigma^T$ changes with increasing food web complexity as well as changes in chemical versus electromagnetic potentials.

## 3. Results

We investigate three aspects of the MEP-optimized trait-based model described above, which include 1) how the three phytoplankton temporal strategies alter the food webs ability to dissipate electromagnetic and chemical potentials, 2) how food web complexity changes entropy production and 3) how the food web adjusts to changes in the relative inputs of electromagnetic versus chemical potential. All simulations are run for two years that include both diel and seasonal changes in solar input but constant inputs of chemical constituents at the specified dilution rate. Only simple food webs are explored in this study, consisting of 1 phytoplankton, 1 bacterium and 1 consumer, $1P1B1C$, and two other configurations consisting of $2P2B2C$ and $3P3B3C$.

*3.1 Phytoplankton temporal strategies and entropy production*

As discussed in Sections 2.6 and S3.2, the three phytoplankton temporal strategies we explore here consist of a balanced growth strategy with no change in phytoplankton stoichiometry, a passive strategy where phytoplankton can store carbon fixed from photosynthesis for later use, and an explicit clock-based, or circadian, strategy for resource allocation between carbon fixation, $r_{1,P\{i\}}$, and biosynthesis, $r_{2,P\{i\}}$, reactions on a diel cycle. However, before presenting the results from those simulations, it is useful to characterize the magnitude of energy inputs for the nominal simulations as well as examine entropy production and phytoplankton dynamics for randomly selected trait values.

The nominal simulations examine a $1P1B1C$ food web model operated at a dilution rate of 0.2 d$^{-1}$ with input concentrations given in Table 2. Under these input conditions, aerobic oxidation of all the supplied organic carbon ($C_L + C_D$) in a 1 m deep pond with 1 m$^2$ surface over a two-year period produces 0.025 MJ K$^{-1}$ of entropy from chemical energy dissipation, while the input of solar radiation from a latitude of 42° over the 1 m$^2$ surface produces 27.1 MJ K$^{-1}$ of entropy over the same two year period. Consequently, energy input and entropy production from electromagnetic radiation is more than a 1000 times greater than that from chemical potential in the nominal simulations. In Section 3.3, results from simulations where energy potentials are similar will be examined, but in this section and the next, nominal conditions (Table 2) will be used that vastly favor dissipation of electromagnetic radiation.

Table 2. Concentrations of state variables in the feed, as well as environmental conditions for the nominal simulations, where $I_s$ is ionic strength and $I_0^M$ is the maximum surface solar radiation at 0° latitude.

| Input | Value | Input | Value |
|---|---|---|---|
| $I_0^M \ [\![mmol\text{-}\gamma\ m^{-2}\ d^{-1}]\!]$ | 406,000 | $[C_L]\ [\![mmol\ m^{-3}]\!]$ | 10 |
| $T\ [\![K]\!]$ | 293 | $[C_D]\ [\![mmol\ m^{-3}]\!]$ | 100 |
| $pH$ | 8.1 | $[N_D]\ [\![mmol\ m^{-3}]\!]$ | 7 |
| $I_s\ [\![M]\!]$ | 0.72 | $[\mathfrak{S}_{P\{i\}}]\ [\![mmol\ m^{-3}]\!]$ | 0.1 |
| $[DIC]\ [\![mmol\ m^{-3}]\!]$ | 2,000 | $[C_{P\{i\}}]\ [\![mmol\ m^{-3}]\!]$ | 0.1 |
| $[O_2]\ [\![mmol\ m^{-3}]\!]$ | 225 | $[\mathfrak{S}_{B\{i\}}]\ [\![mmol\ m^{-3}]\!]$ | 0.1 |
| $[NH_3]\ [\![mmol\ m^{-3}]\!]$ | 5 | $[\mathfrak{S}_{C\{i\}}]\ [\![mmol\ m^{-3}]\!]$ | 0.1 |

The number of traits, $n_T$, that need to be set depends on the size of the food web and is given by, $4n_P + 3n_B + (1 + n_P + n_B + n_C)n_C$; consequently, the $1P1B1C$ has an 11 dimensional trait space. To examine entropy production and food web dynamics when traits are randomly assigned, we conducted 90 simulations where the trait values were randomly selected based on Latin hypercube sampling. In these simulations, total entropy production, $\sigma^T$, over a two-year period varied from 0.2994 to 11.91 MJ K$^{-1}$, and phytoplankton reached a maximum concentration of approximately 35 mmol m$^{-3}$ with various diel and seasonal dynamics (Figure 2). Furthermore, if the growth efficiencies for phytoplankton, bacteria and consumers are set to zero so that no growth occurs, the total entropy produced is 0.3077 MJ K$^{-1}$ due to particle absorption of solar radiation by biomass in the input (Table 2).

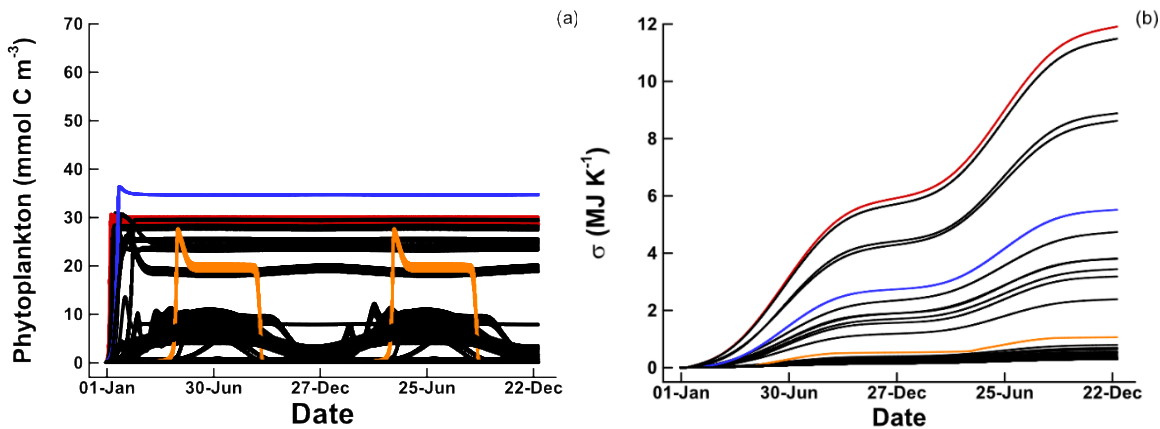

Figure 2. (a) Phytoplankton and (b) cumulative total entropy production, $\sigma^T$, over two years for 90 simulations with random selection of trait values for a $1P1B1C$ food web configuration using nominal input concentrations (Table 2). Colored lines in (a) and (b) correspond to the same solutions.

For each temporal strategy, optimizations were run starting from 90 different initial locations within trait space that were selected by Latin hypercube sampling (see Section S3) to increase the likelihood of locating the global optimum. All three temporal strategies produced similar phytoplankton concentrations between 50 and 60 mmol m$^{-3}$ (Figure 3a) and all reduced NH$_3$ ammonium concentrations from the input of 5 mmol m$^{-3}$ to 0.5 to 2 mmol m$^{-3}$ (Figure 3d). Similarly, recalcitrant nitrogen, $N_D$, was drawn down from 7 mmol m$^{-3}$ to approximately 1.5, 1.2, and 0.7 mmol m$^{-3}$ in the balanced growth, passive and circadian simulations, respectively. In all three simulations, optimal solutions prevented consumers from growing by selecting consumer growth efficiencies, $\varepsilon_{C\{1\}}$, near 0 or 1 (Table 3). Bacteria concentrations were highest in the circadian strategy, at 12 mmol m$^{-3}$, and lowest in the passive strategy, at 2.5 mmol m$^{-3}$ (Figure 3b). Both the circadian and passive strategies accumulated high concentrations of phytoplankton internal carbon, $C_{P\{1\}}$, with a strong seasonal signal, while $C_{P\{1\}}$ in the balanced growth strategy never exceeded 70 mmol m$^{-3}$ (Figure 3c). The high $C_{P\{1\}}$ concentrations in the passive and circadian strategies produced phytoplankton C:N ratios that varied from 115 to 230, and from 140 to 270, respectively, while the phytoplankton C:N ratio in the balanced growth solution was held fixed at 14 (data not shown).

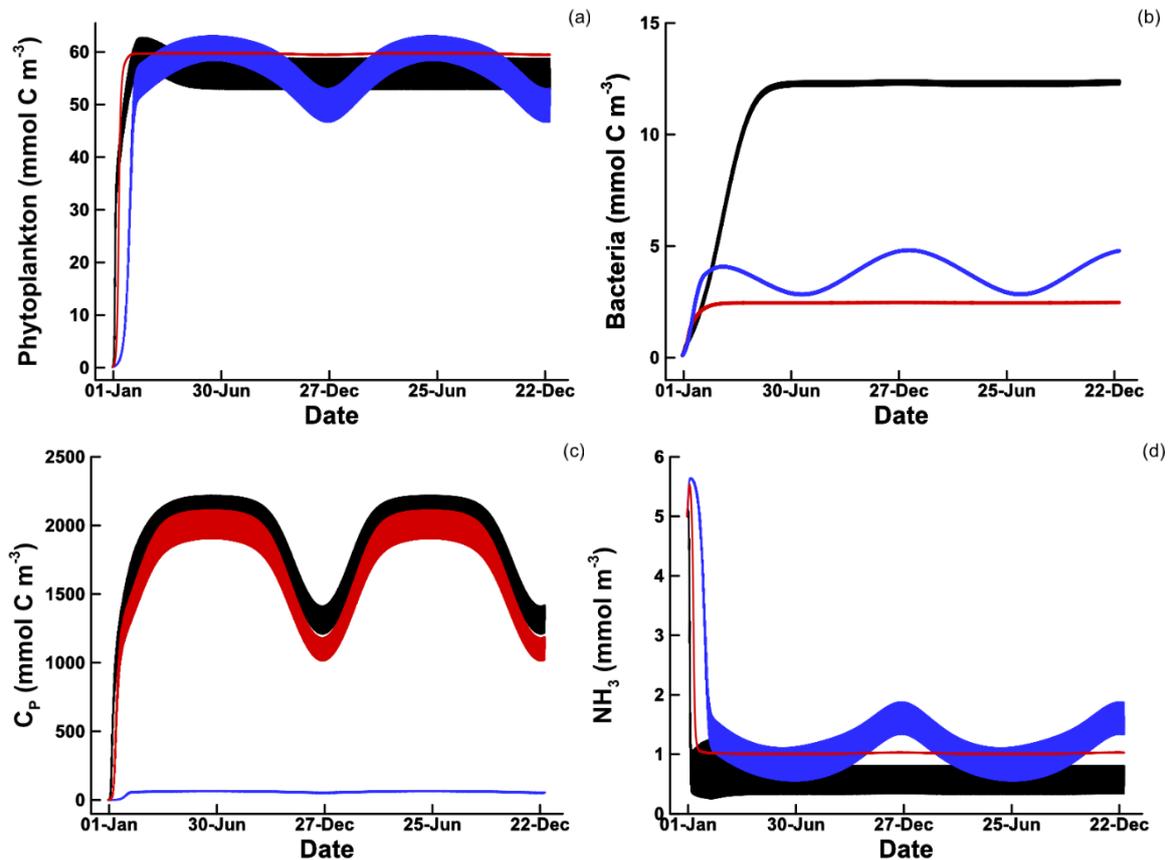

Figure 3. Variations in (a) phytoplankton, (b), bacteria, (c) phytoplankton carbon storage $C_P$, and (d) ammonium concentrations (mmol m$^{-3}$) over the two-year simulations under the three different temporal strategies: Blue, balanced growth; Red, passive storage; Black, circadian allocation.

Table 3. Optimal trait values obtained from maximizing total entropy production, $\sigma^T$, in a $1P1B1C$ food web model over a two-year period for the three different temporal strategies under nominal conditions (Table 2) at a dilution rate of 0.2 d$^{-1}$.

| Variable | Balanced Growth | Passive Storage | Circadian Allocation |
|---|---|---|---|
| $\varepsilon_{P\{1\}}$ | 0.2536 | 0.3452 | 0.3788 |
| $t_{On\{1\}}$ (d) | 0.0000* | 0.0000* | 0.2389 |
| $t_{Off\{1\}}$ (d) | 1.0000* | 1.0000* | 0.7799 |
| $\Omega_{amp\{1\}}$ | 0.6746 | 0.8036 | 1.0000 |
| $\varepsilon_{B\{1\}}$ | 0.1686 | 0.1618 | 0.1628 |
| $\Omega_{1,B\{1\}}$ | 0.3351 | 0.3852 | 0.4349 |
| $\Omega_{2,B\{1\}}$ | 0.1997 | 0.1609 | 0.2869 |
| $\Omega_{3,B\{1\}}$ | 0.4651 | 0.4539 | 0.2782 |
| $\varepsilon_{C\{1\}}$ | 0.0001 | 0.9971 | 0.0001 |
| $\Omega_{P\{1\},C\{1\}}$ | 0.7288 | 0.0000 | 0.6547 |
| $\Omega_{B\{1\},C\{1\}}$ | 0.0729 | 0.3888 | 0.5809 |
| $\Omega_{C\{1\},C\{1\}}$ | 0.4396 | 0.5868 | 0.0963 |
| $\sigma^R$ (MJ K$^{-1}$) | 0.2826 | 0.9901 | 0.9787 |
| $\sigma^W$ (MJ K$^{-1}$) | 0.2751 | 1.687 | 1.616 |
| $\sigma^P$ (MJ K$^{-1}$) | 3.426 | 17.79 | 18.60 |
| $\sigma^T$ (MJ K$^{-1}$) | 3.984 | 18.95 | 19.74 |

* These values were held constant and not part of optimization.

In terms of entropy production over the two year simulation, the circadian strategy produced the greatest amount at 19.74 MJ K$^{-1}$, followed closely by the passive strategy at 18.95 MJ K$^{-1}$, which where both much higher than the balanced growth solution, which only produced 3.984 MJ K$^{-1}$ (Table 3). The source of entropy production for all three strategies is largely due to light attenuation due to particles in the water column, with $\sigma^P$ accounting for 86%, 94% and 94% for the balanced growth, passive storage and circadian strategies, while entropy production from reactions, $\sigma^R$, only accounts for 7.1%, 5.2%, and 5.0% of $\sigma^T$, respectively (Table 3). These results are not surprising given the amount of energy entering the system from light versus chemical potential. For aquatic systems, dissipating electromagnetic potential is mostly about synthesizing particles to intercept and dissipate as heat high frequency photons [17]. Light attenuation by water contributes 6.9%, 0.89% and 0.82% to entropy production, $\sigma^W$, for the balanced, passive and circadian strategies, respectively.

The entropy production differences between the three strategies can be understood by considering how phytoplankton biomass is allocated to the carbon fixation reaction, $r_{1,P\{1\}}$, and the biosynthesis reaction, $r_{2,P\{1\}}$, that is determined by $\Omega_{1,P\{1\}}$, which in turn depends on the three trait values, $t_{On\{1\}}$, $t_{Off\{1\}}$, and $\Omega_{amp\{1\}}$, that govern the nature of the square wave function, Eqn (S76). For both the balanced growth and passive storage strategies, $t_{On\{1\}}$ and $t_{Off\{1\}}$ are constrained to be 0 and 1, respectively, so that $\Omega_{1,P\{1\}}$ remains constant at the value given by $\Omega_{amp\{1\}}$, while $t_{On\{1\}}$ and $t_{Off\{2\}}$ can be any value in the range [0,1] for the circadian strategy, which produces a diel square wave when either $t_{On\{1\}} > 0$ or $t_{Off\{1\}} < 1$ (Figure 4a). In the optimal circadian strategy, all phytoplankton resources are allocated to CO$_2$ fixation (Table 3; $\Omega_{amp\{1\}} = 1$) when the fractional time of day, $t_D$, falls within the interval $0.2389\,d \leq t_D \leq 0.7799\,d$, and redirected to biosynthesis outside the interval (Figure 4a, black lines). In both the balanced growth and passive storage strategies, the optimal solutions locate a compromise between allocation of biomass to $r_{1,P\{1\}}$ and $r_{2,P\{1\}}$, where 67.5% and 80.4% of biomass is allocated to carbon fixation at all times for the passive storage and balanced growth optimal solutions, respectively ($\Omega_{amp\{1\}}$, Table 3; Figure 4a, blue and red lines). These different allocation strategies significantly impact the rates of the two reactions associated with phytoplankton ($r_{1,P\{1\}}$ and $r_{2,P\{1\}}$).

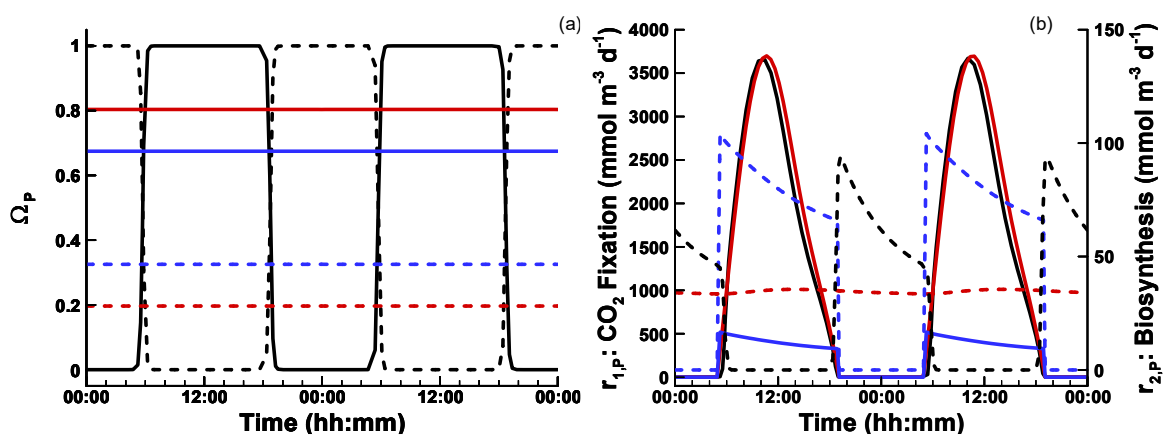

Figure 4. (a) How phytoplanton resource allocation, $\Omega_{1,P\{1\}}$ (solid lines) and $\Omega_{2,P\{1\}}$ (dashed lines) and (b) reactions for CO$_2$ fixation, $r_{1,P\{1\}}$, (solid lines) and biosynthesis, $r_{2,P\{1\}}$ (dashed lines), vary over a two day period in the two year simulations associated with balanced growth (blue lines), passive storage (red lines) and circadian resource allocation (black lines).

In the balanced growth strategy, $r_{1,P\{1\}}$ and $r_{2,P\{1\}}$ must be coupled by definition of balanced growth; consequently, growth can only occur during the day, and any limitations on growth in the day, such as due to $NH_3$ available, also limit $CO_2$ fixation rate. For instance, considering just two days in the two year simulation (Figure 4b), $r_{1,P\{1\}}$ and $r_{2,P\{1\}}$ in the balanced growth strategy (Figure 4b, blue lines), both equal 0 at night, and the decrease in both $CO_2$ fixation (blue solid line) and biosynthesis (blue dashed line) during the day is due to $NH_3$ limitation occurring (Figure 4b). The passive storage strategy avoids the reaction coupling limitation, so that biosynthesis can occur at night (Figure 4b, dashed red line), but because resource allocation is fixed in the passive strategy, all cellular resources cannot be allocated to growth at night. The circadian strategy relaxes this problem by allocating resources dynamically, so that growth at night can be maximized (Figure 4b, black dashed line), yet still be able to fix $CO_2$ during the day at maximum rate as well (Figure 4b, black solid line). However, since chemical reactions contribute little to entropy production in all strategies (Table 3, $\sigma^R$), the source of entropy production lies elsewhere.

The much higher entropy production by the passive and circadian strategies over the balanced growth strategy is due to the much higher concentration of light absorbing particles. While phytoplankton concentrations are similar in all three strategies (Figure 3a), phytoplankton carbon storage, $C_P$, that contributes to light attenuation (Eqn. (S8)) is much higher in the passive and circadian strategies (Figure 3c). In order to store carbon, phytoplankton must grow in size, which increases their cross-sectional area for light interception. The intercepted light cannot be used for photosynthesis by definition, but it is converted to heat that results in entropy production. Under the nominal conditions (Table 2), the planktonic community is N limited, so biomass, $S$, accumulation is constrained; however, $C_P$ does not contain N and is not constrained by N availability. Both the passive and circadian strategies increase entropy production by investing in $C_P$ synthesis using electromagnetic radiation, and the circadian strategy does this slightly more effectively due to temporal control on resource allocation. Not surprisingly, entropy production in the balanced growth solution is increased by increasing either $NH_3$ or $N_D$ concentrations in the feed. For instance, at an $N_D$ input concentration of 50 mmol m$^{-3}$, the balanced growth solution increases $\sigma^T$ to 15.33 MJ K$^{-1}$ and maintains a phytoplankton concentration of ~300 mmol m$^{-3}$.

*Entropy production and food web complexity*

Here we examine the effect of increasing food web complexity by adding more ecotypes to each functional group. In particular, we compare the solutions from the $1P1B1C$ configuration discussed above to $2P2B2C$ and $3P3B3C$ configurations run under the nominal input conditions (Table 2). These simulations, which were also run at a dilution rate of 0.2 d$^{-1}$, produced nearly the same amount of entropy as the $1P1B1C$ solution (Table 4), and nutrient and organism dynamics were very similar to the $1P1B1C$ solutions as well, with only minor or duplicate contributions from the additional ecotypes (data not shown). For instance, in the $3P3B3C$ food web using passive storage strategy, two phytoplankton exhibited nearly identical dynamics and each attained a steady state concentration of ~30 mmol m$^{-3}$, so when summed together they were equivalent to the $1P1B1C$ solution (Figure 3a, red line). As in the $1P1B1C$ solutions, consumers were nearly absent. The additional ecotypes in the more complex food web were effectively superfluous as far as the entropy maximization is concerned. However, the complexity of the food web become more important as dilution rate was increased, as well as the circadian strategy compared to the passive strategy.

Table 4. Total entropy production, $\sigma^T$, from three different food web configurations over a two-year period with the three different temporal strategies under nominal conditions (Table 2) at a dilution rate of 0.2 d$^{-1}$.

| Strategy | $1P1B1C$ | $2P2B2C$ | $3P3B3C$ |
|---|---|---|---|
| **Balanced** | 3.9837 | 4.034 | 4.0811 |
| **Passive** | 18.9511 | 18.9976 | 19.0152 |
| **Circadian** | 19.7359 | 19.7844 | 19.7987 |

When the three strategies along with the three different food webs were run under nominal input concentrations but at a dilution rate of 1.5 d$^{-1}$, the added food web complexity and the circadian strategy showed enhanced entropy production relative to the other simulations (Table 5). There is approximately a 7% to 12% increase in $\sigma^T$ associated with the increase in food web complexity from $1P1B1C$ to $2P2B2C$ or from $2P2B2C$ to $3P3B3C$ regardless of the temporal strategy employed, but a much greater increase in $\sigma^T$ occurred as temporal strategies were changed. There is approximately a 320% increase in $\sigma^T$ as the strategy was changed from balanced growth to passive storage. In fact, phytoplankton in the solutions using the balanced growth strategy wer near washout conditions at a dilution rate of 1.5 d$^{-1}$, as their concentrations only attain ~1 mmol m$^{-3}$ for a short period during the peak of summer. When the temporal strategy was switched from passive to circadian, there was approximately a 130% increase in $\sigma^T$, which indicates the usefulness of an explicit clock in improving entropy production over the passive solution. Furthermore, optimal solutions at high dilution rates exhibited complementary when more complex food webs were used.

Table 5. Total entropy production, $\sigma^T$, from three different food web configurations over a two-year period, each run using the three different temporal strategies under nominal concentration (Table 2) but at a dilution rate of 1.5 d$^{-1}$.

| Strategy | $1P1B1C$ | $2P2B2C$ | $3P3B3C$ |
| --- | --- | --- | --- |
| Balanced | 0.3226 | 0.3455 | 0.3684 |
| Passive | 1.3374 | 1.4995 | 1.6432 |
| Circadian | 3.0891 | 3.4000 | 3.6242 |

In general, most of the complex food web (2x and 3x) solutions did not show much complementarity between ecotypes at low dilution rates; however, when dilution rate was increased to 1.5 d$^{-1}$ or more, phytoplankton (as well as the other functional groups to a lesser extent) exhibited complementary in solutions using the $2P2B2C$ or $3P3B3C$ food webs with either the passive storage or circadian strategies (Figure 5). For instance, at a dilution rate of 1.5 d$^{-1}$, the best circadian solution select for phytoplankton with traits that are complementary with respect to winter versus summer (Figure 5a, red versus black lines). (Note, simulations did not investigate seasonal fluctuations in temperature, just solar radiation.) In the circadian solution, $\varepsilon_{P\{1\}}$ and $\varepsilon_{P\{2\}}$ equal 0.341 and 0.401, respectively, which allows $\mathcal{S}_{P\{2\}}$ to grow slightly more efficiently than $\mathcal{S}_{P\{1\}}$ giving the former an advantage during winter when light intensity is lower and NH$_3$ is a ~2 mmol m$^{-3}$ higher. The advantage of the circadian strategy over the passive strategy is evident in phytoplankton concentrations between the two simulations. At the same dilution rate of 1.5 d$^{-1}$, the passive strategy has lower summer time phytoplankton concentration, ($P\{2\}$, Figure 5b, black line), and the winter ecotype, $P\{1\}$, is closer to being washed out of the system (Figure 5b, red line), which results in less entropy production compared to the circadian strategy (Table 5). At a dilution rate of 2.0 d$^{-1}$, the $2P2B2C$ food web using the circadian strategy has an entropy production of 1.5042 MJ K$^{-1}$ and looks very similar to Figure 5b, which implies the circadian strategy has approximately a 0.5 d$^{-1}$ specific growth rate advantage over the passive strategy, which only produces 0.5943 MJ K$^{-1}$ at the 2.0 d$^{-1}$ dilution rate.

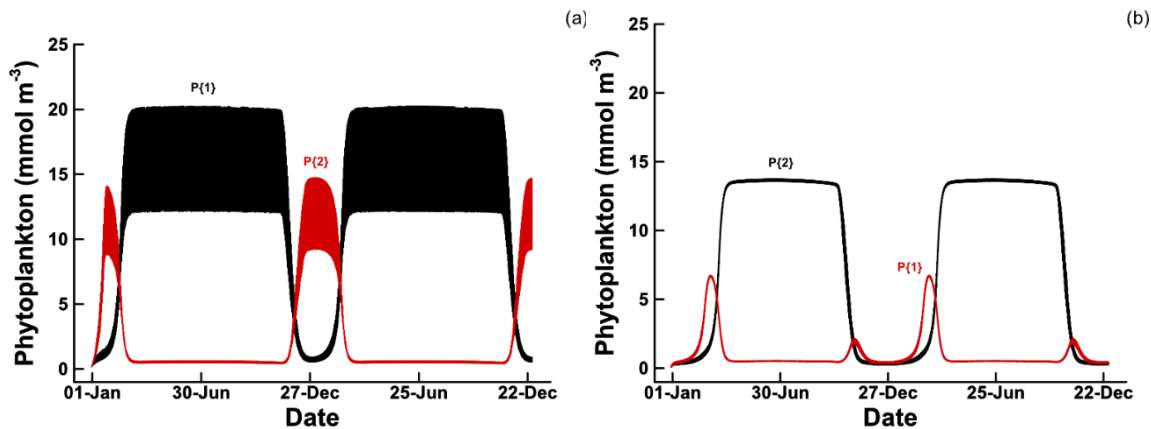

Figure 5. Phytoplankton concentration for a $2P2B2C$ food web configuration at a dilution rate of 1.5 d$^{-1}$ using (a) the circadian allocation strategy versus (b) the passive storage strategy.

*3.3 Dissipation of chemical versus electromagnetic potential energies*

As mentioned earlier, dissipation of all electromagnetic potential produces more than 1000 times the entropy than dissipation of the chemical potential under nominal inputs (Table 2). In this section we examine simulations where the electromagnetic potential is reduced by a factor of 10 ($I_0^M = 40,600.\ (mmol\text{-}\gamma\ m^{-2}\ d^{-1})$), and the chemical potential is increased by a factor of 100 by increasing the input concentration of $C_L$ from 10 mmol m$^{-3}$ to 12 mol m$^{-3}$. With these changes, maximum possible entropy produced over a 2-year period from electromagnetic radiation reduces to 2.7068 MJ K$^{-1}$ and that from chemical potential increases to 2.8061 MJ K$^{-1}$. Simulations are run at a dilution rate of 0.2 d$^{-1}$ with a $1P1B1C$ food web configuration using only the circadian temporal strategy, and we only consider a single optimization run using the standard 90 initial conditions in trait space based on Latin hypercube sampling.

Inspection of all 90 simulations reveals what appears to be three different types of solutions based on entropy production (Figure 6a). The first 36 solutions all produce nearly the same amount of entropy over the two year period of 0.3019 MJ K$^{-1}$ (Figure 6a, red lines), and all these solutions invest mainly in bacteria (Figure 6b, red lines) to oxidize $C_L$ from the initial 12.00 down to 11.232 mol m$^{-3}$ but leave $C_D$ unused. No phytoplankton are produced (Figure 6c, red lines), and for most solutions, consumers remain at low concentrations (Figure 6d, red line). The next 16 solutions locate an entropy maximum that is a little higher at an average of 0.4583 MJ K$^{-1}$ (Figure 6a, blue lines). These solutions do not invest in phytoplankton either and still produce entropy by bacterial oxidation of $C_L$, but these solutions lower the concentration of $C_L$ further to 10.485 mol m$^{-3}$ by investing some N resources in consumers, which results in lower bacteria concentrations (Figures 6b and 6d, blue lines). By investing in consumers, which remineralize N in bacteria as NH$_3$ and $N_D$ by grazing, the strategy reduces the N limitation on bacterial growth by rapid recycling allowing them to consume more $C_L$ and produce more entropy than solutions without consumers.

The last 38 solutions instead invest in phytoplankton (Figure 6c, grey and black lines) to tap the electromagnetic potential, producing the greatest amount of entropy at 0.9041 MJ K$^{-1}$ and imparting smooth oscillations in cumulative entropy production due to the seasonal nature of solar radiation over the two year period (Figure 6a, grey and black lines). There appears to be either several local optimum in these solutions, or the optimization routine may have had difficulty locating the true global optima because the entropy production from the 38 solutions span a range from a minimum of 0.4846 to the maximum of 0.9041 MJ K$^{-1}$ (Figure 6a, grey and black lines). These solutions invest minimally in bacteria (Figure 6b), which are used primarily to remineralize $N_D$ to NH$_3$, which is evident in values of $\Omega_{1,B\{1\}}$, $\Omega_{2,B\{1\}}$ and $\Omega_{3,B\{1\}}$ traits. In the first 52 of 90 solutions discussed above, the reaction for bacterial growth, $r_{1,B\{1\}}$, is heavily favored with $\Omega_{1,B\{1\}} \cong 0.95$, with the remainder of the bacterial catalyst allocated to $N_D$ decomposition by $r_{3,B\{1\}}$ with $\Omega_{3,B\{1\}}$ set to 0.05. In the phytoplankton-based strategy, the weighting of bacterial catalyst to reactions is more variable, but solutions are in the nationhood defined by $\Omega_{1,B\{1\}} \cong 0.5$ and $\Omega_{3,B\{1\}} \cong 0.5$. In all solutions $\Omega_{2,B\{1\}} \cong 0$, so that $C_D$ remains unused.

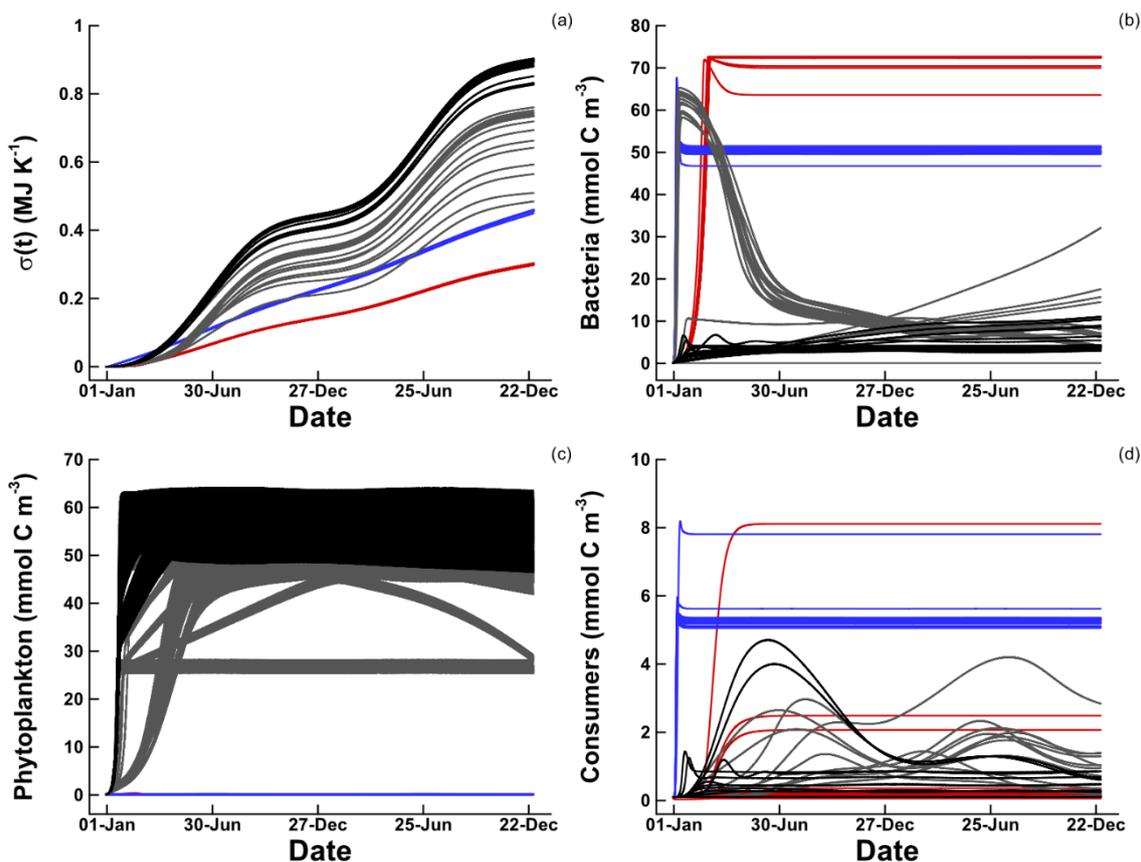

Figure 6. All 90 solutions from simulations using a $1P1B1C$ food web configuration with phytoplankton circadian allocation strategy where the input $C_L$ concentration has been increased to 12 mole m$^{-3}$ and the solar radiation has been decreased by a factor of 10 so that electromagnetic and chemical potentials are nearly equal. (a) Cumulative total entropy production, (b) bacteria, (c) phytoplankton and (d) consumer concentrations over the two year simulation, where the line colors highlight solutions grouped around the three different optimum attractors corresponding to bacteria only (red), bacteria plus consumers (blue) and phytoplankton (grey+black).

## 4. Discussion

One of the challenges in modeling biogeochemistry is that numerous parameters are needed to describe growth kinetics and predator-prey interactions of the organisms that comprise a microbial food web [47]. Because natural microbiomes consist of hundreds to thousands of species whose growth kinetics and interactions are poorly known, models typical aggregate organisms into functional groups, such as the three used in this study. Even after aggregation, many dozens of parameters remain, and their values are unknown and only crudely bounded. Consequently, parameter values are "tuned" so that the sum of the squared residuals between model output and observations is minimized [48-50]. Because the models contain little to no fundamental principles other than conservation of mass, the standard approach becomes a non-linear modeling fitting exercise, and there is often multiple parameter values that can fit the limited observations equally well [48]. While the resulting models do interpolate observations well, their ability to predict beyond the observations used for calibration is very limited, because fundamental information on how microbial communities organize and function is lacking in their development. Furthermore, the models require recalibration as environmental conditions change even for interpolation, because new conditions drive species succession that have different growth characteristics.

To address some of the deficiencies of classic food web models, there is a long history of applying thermodynamic approaches to understand ecosystem function that forego predicting fine-scale details to improve long-term prediction fidelity, which is similar to predicting climate rather than weather [5,51]. More recently, trait-based modeling has been developed that avoids the brittleness of classic models by retaining high species diversity that allows *in silico* community succession. The current formulation of the thermodynamically-based MEP model gains inspiration from trait-based models [43] that have advanced to include size-structured food webs [52] as well as functional gene expression [53] and adaptation [54]. However, we have proposed a new approach where traits are not randomly assigned but rather are determined from maximizing entropy production, which allows fewer ecotypes to be simulated thereby reducing model state dimension. While the MEP trait-based model does require a computationally costly optimization, once trait values are determined, the model can be run without optimization as we will discuss at the end of this section.

The model presented in this manuscript is designed under the hypothesis that systems organize to dissipate energy potentials, and specifically that living systems maximize entropy production over a characteristic time scale while abiotic systems maximize entropy production instantaneously. Interestingly, the objective function is not growth oriented, as that alone does not dissipate free energy, but rather to produce catalysts that dissipate chemical potentials or produce particles that intercept light and dissipate it as heat. Unlike most bio-centric models that strive to grow organisms and/or maximize growth rate, the emphasis here is on energy potentials and how the system can organize to maximize their rate of destruction. As Lineweaver and Egan [55] noted, 'This represents a paradigm shift from "we eat food" to "food has produced us to eat it".' Furthermore, we have placed emphasis on removing biological parameters and replacing them with control variables, or traits, which are dynamically adjusted between model runs to maximize EP over a specified time interval. If the model does not produce realistic results or compares poorly to observations, it indicates that either model structural errors exist or that the MEP hypothesis is falsified for biogeochemistry [23]. The MEP approach also permits quantitative comparison of different energy sources to a system, such as our comparison of chemical versus electromagnetic potentials.

This manuscript's primary focus is on how temporal strategies increases entropy production over time-agnostic strategies. Temporal strategies are one of the hallmarks of biology, such as circadian clocks [56], anticipatory control [33,34], energy and resource storage [31,32], dormancy and persister cells [35], and resource time sharing [57]. More recently, microbial communities have been found to exhibit circadian dynamics as well [30,58,59], but other than passive storage [44,60], marine biogeochemistry models typically do not include such mechanisms. Our results show that including temporal strategies results in significantly greater entropy production than balanced growth, and that explicit strategies, such as the simple diel square-wave function, increase entropy production further over passive strategies, especially near the upper limits of phytoplankton growth. These results are consistent with observations that show how bacteria shift cellular metabolic function to cope with fluctuating environments [61] and how phytoplankton and other members of the community change metabolic expression over diel cycles [29,58]{Tsakalakis, 2018 #6604}. There are systems biology models that have been constructed for phytoplankton that show the importance of time-dependent resource allocation [62], but these are not at an ecosystems scale, which is the focus of our study.

The MEP model using the circadian temporal strategy also makes some predictions regarding specific growth rate and C:N composition of phytoplankton, although we did not explore these areas in detail. The chemostat simulations indicate a maximum phytoplankton growth rate around 2 d$^{-1}$ before washout occurs, which is near observed maximum phytoplankton growth rates at 20°C [63]. These results are encouraging in that our formulation does not include a parameter for maximum specific growth rate like standard kinetic models, but only uses photon interception rate combined with an MEP-determined trait on growth efficiency, $\varepsilon_{P\{i\}}$, to set the specific growth rate. We have not included temperature dependency in the MEP model, so we cannot compare our results to the full Eppley curve [64], but adding temperature dependency would be worth developing. The addition of the phytoplankton carbon storage improves growth rate for both the passive and circadian strategies over the balanced growth strategy, which is consistent with observation on improving competitive advantage in fluctuating environments [62,65].

The model also predicts phytoplankton C:N ratio, which varies between 140:1 and 270:1 for the nominal simulation, which is considerably higher than the Redfield ratio of 6.6:1. The C:N ratio for phytoplankton is known to vary as a function of growth rate and nutrient limitations, but the maximum observed values are closer to 50:1 [66-68]. In the MEP simulations, the high levels of internal carbon storage in both the passive and circadian strategies is used to capture and dissipate light to enhance entropy production, which differs from how storage is typically used, such a survival in fluctuating environments [65]. The internal C storage, $C_{P\{i\}}$, behaves as a particle in the light attenuation model (Eqn (S8)); consequently, one mechanism to dissipate electromagnetic radiation when N is limiting is to increase $C_{P\{i\}}$ concentration, which is what the passive and circadian strategies implement since there is no model constraint on the C:N ratio of phytoplankton. Effectively, the high phytoplankton C:N ratio is a consequence of the light attenuation model used. As far as we know, there have not been studies that examine how light attenuation changes with phytoplankton internal carbon stores, so our first order approximation assumed a linear relationship between the light attenuation factor, $k_{wp}$, and the concentration of phytoplankton carbon storage, $[C_{P\{i\}}]$. Since a linear formulation results in excessively high phytoplankton C:N ratios, that assumption should be revisited in future versions of the MEP model, which we discuss at the end of this section.

Simulations with the more complex food webs, $2P2B2C$ and $3P3B3C$, did not show much improvement in entropy production when specific growth rates where low (0.2 d$^{-1}$), but they did produce more entropy at higher specific growth rates (1.5 and 2.0 d$^{-1}$) compared to the $1P1B1C$ food web configuration. At the higher specific growth rates, the more complex models discovered complementarity [69,70], where trait values for one phytoplankton specialized in high light intensity during the summer and another ecotype had trait values that performed better under winter conditions. Complementarity is also exhibited in trait-based models [71] provided the initial population contains ecotypes with diverse parameterizations. In our approach, this is not necessary as the optimization sets the trait values and will select for complementary ecotypes when the strategy increases entropy production.

The optimization approach also provides a potential solution to food web closure. In standard compartment models as well as trait-based models, there is often uncertainly on the number of trophic levels that should be included in the model, and it has been demonstrated that the type of closure, which refers to a top predator that is not formally included as a state variable in the model, dramatically changes simulation results [72-74]. In most natural environments there is usually some limiting resource, such as energy input or chemical constituent. From an optimization perspective, how should the limiting resource, such as N, be allocated to phytoplankton versus bacteria versus consumers? In our formulation, the structure of the trophic levels described by the matrix $\Omega_{\chi\{j\},C\{i\}}$ is determined as part of the optimization. In section 3.3, the optimizations located three general attractors, the first being a solution that did not allocate much resources to the consumers (bacteria only), while the second optimum attractor did allocate resources to consumers, which resulted in greater entropy production. While often underappreciated, predators can increase the growth rate of their prey by increasing recycling of nutrients the prey is limited by [75-78]. In essence, the addition of predators increase the rate at which a system cycles, and since the cycle is powered by energy dissipation, the presence of predators can lead to increase entropy production, as was found by the second optimum attractor in our study. Predation can enhance entropy production.

Casting the model in the currency of entropy production allows comparison between abiotic and biotic processes as well as comparison of different free energy sources driving system organization. The input concentrations in the nominal simulations reflect values found in natural systems, but by casting the model in entropy production, or energy input, it revealed that solar radiation dominated energy input compared to chemical potential by more than a thousand fold. When the model was rerun with near equal inputs of electromagnetic and chemical free energy, three different attractors where identified consisting of bacteria only, bacteria with consumer predation and phytoplankton only. While the latter solution was found to be the global maximum, the other solutions were locally stable. Furthermore, while we mostly presented only the best solution from the 90 optimizations run, other solutions that were near the maximum entropy production were also found, and some of these solutions exhibited different dynamics of the functional groups or chemical species. In fact, one of the requirements for systems to follow MEP trajectories is that there must be multiple degrees of freedom in the system, and that there are often many solutions that can generate equivalent entropy production [3]. Examining local optima revealed by the MEP model may shed light on ecosystem stability and tipping points, as we would expect an ecosystem to shift over time to higher entropy producing states, especially if new states arise due to environmental change [79-81].

We end this section with a few research directions for entropy guided trait-based modeling. All of the simulations explored in this manuscript examined steady state inputs, except for solar radiation that included diel and season variations. Because of the stability of the inputs, complex food webs involving many ecotypes of each function group did not provide much improvement in entropy production in low growth scenarios; however, if simulations were run with time varying inputs or step changes, we would expect the higher number of environment niches would drive optimal solutions to exhibit complementarity. For instance, solutions would likely include ecotypes with oligotrophic or copiotrophic growth kinetics, or high light versus low light ecotypes, if those niches were present during the simulation. Consequently, it might be possible to conduct trait-based optimization in 0D to develop food webs capable of high entropy production under a number of different environmental conditions. Because the optimizations are computationally expensive, conducting them in 0D environments would greatly increase speed of determining trait values for optimum food webs. Once determined, the optimized food web could be run in 3D global circulation models without the computationally costly optimization component.

Another area that needs advancing is the light attenuation model. We only examined blue light, and the light attenuation model (Eqn. (S8)) is rather simplistic. It is known that phytoplankton can significantly change their light attenuation characteristics of Chlorophyll by at least an order of magnitude, and attenuation characteristics vary as a function of wavelength as well [42]. It seems likely that there are energy and resource use tradeoffs in synthesizing different types of light harvesting compounds, but those relationships are not well known. Consequently, exploring resource allocation and light harvesting is needed, since electromagnetic potentials are the dominant energy input to many ecosystems.

Using information for genome scale metabolic network models [82] could also be useful in defining the reactions used the distributed metabolic network for the trait-based model, and the reaction governing how consumers remineralize resources as labile versus recalcitrant (Eqn. (S50)) is in need of further research. Perhaps the most interesting question, though, is what are the time scales over which biological strategies operate? What temporal strategies has biology learned over 3.5 billion years of evolution to facilitate entropy production under the guise of Darwinian growth? Comparing MEP-based simulations to experiments and observations may be one means of answering these questions.

## 5. Conclusions

Under the hypothesis that biological systems organize to dissipate energy potentials over a characteristic time scale, we have investigated how three temporal strategies affect entropy production in a simple marine food web model consisting of phytoplankton, bacteria and consumer functional groups. The balanced growth strategy, where phytoplankton grow with fixed stoichiometry, was found to produce the least amount of entropy because growth can only occur when light is present. A significant increase in entropy production occurred with a passive storage strategy that allowed phytoplankton to accumulate reduced carbon during the day to fuel phytoplankton growth at night. The best solution, however, was attained by including an explicit circadian clock that dynamically allocated resources to energy harvesting versus biosynthesis reactions to optimally use diel input of solar radiation. We also demonstrated a new type of trait-based modeling that used entropy production maximization to determine trait values as opposed to the standard method of allowing *in silico* natural selection to cull the population of poor performers.

Our results illustrate that organisms that evolve the ability to predict future conditions via explicit temporal strategies can increase entropy production. The time scale over which biological systems have evolved to operate, however, remains an open but important question.

**Supplementary Materials:** Governing equations (S1-S81) with Figure S1 and an example parameter input file.

**Author Contributions:** Conceptualization, JJV and IT; model development, JJV; writing—original draft preparation, JJV and IT; writing—review and editing, JJV and IT; funding acquisition, JJV. All authors have read and agreed to the published version of the manuscript.

**Funding:** This research was funded by the Simons Foundation grant 549941 (JJV, IT) and NSF awards: 1558710 (JJV, IT), 1655552, 1637630, 1841599 (JJV),

**Conflicts of Interest:** The authors declare no conflict of interest.

*Supplementary Material*
# Governing Equations

**S1. Overview**

This *Supplementary Material* describes the details of the model used to demonstrate how temporal strategies can increase entropy production over a given time interval. Constituent transport is governed by a simple well-mixed chemostat of constant volume that receives a constant flow of water with defined input concentrations and is illuminated at the surface with monochromatic light (blue, 440 nm) that varies on both diel and seasonal cycles. The food web consists of three functional groups, phytoplankton, $ⓈⓈ_P$, bacteria, $ⓈⓈ_B$, and consumers, $ⓈⓈ_C$, that produce or consume dissolved inorganic carbon, DIC, oxygen, $O_2$, ammonium, $NH_3$, labile organic carbon, $C_L$, and detrital organic carbon and nitrogen, $C_D$ and $N_D$, respectively (Figure 1). Biological structures for all three functional groups are given the same unit carbon elemental composition, $CH_\alpha O_\beta N_\gamma P_\delta$, but phytoplankton also contain an internal pool of carbon, $C_P$, with elemental composition $CH_2O$. All concentrations are in $[\![mmol\ m^{-3}]\!]$, where double brackets are used to indicate units of variables. The model uses a trait-based approach [1] where each functional group, $ⓈⓈ_{\chi\{i\}}$, is represented by $n_\chi$ ecotypes, or realizations, that are assigned different parameter values that govern reaction stoichiometries, growth kinetics and protein allocation to metabolic pathways, and $\chi$ is $P$, $B$ or $C$. Unlike canonical trait-based models, parameters governing traits (aka control variables [2]) for each ecotype are not randomly assigned but determined by solving a non-linear optimization problem that maximizes integrated entropy production associated with irreversible processes over a fixed simulation period of two years unless otherwise noted.

**S2. Transport, Reaction and Entropy Production Model**

*S2.1 Mass Balance model*

The maximum entropy production (MEP)-optimize trait-based model uses a simple 0D, well-mixed system for transport, where nutrients and low concentrations of organisms flow into a reservoir of volume $V\ [\![m^3]\!]$ at flow rate $F\ [\![m^3\ d^{-1}]\!]$ to produce a dilution rate of $D\ [\![d^{-1}]\!] = \frac{F}{V}$. The pond-like cylindrical reservoir is in contact with the atmosphere at one end, has a cross-sectional area $A\ [\![m^2]\!]$ and depth $d\ [\![m]\!]$ and is illuminated at the surface with photosynthetically active radiation (PAR) of intensity $I_0(t)\ [\![mmol\ photons\ m^{-2}\ d^{-1}]\!]$ that varies both diurnally and seasonally [3]. A simple mass balance around the state variables leads to initial value problem, which, in vector form, is as follows,

$$\frac{d\mathbf{c}(t)}{dt} = D(\mathbf{c}^I - \mathbf{c}(t)) + \frac{A}{V}\mathbf{v}\circ\left(\mathbf{p}\circ\mathbf{h}(T) - \mathbf{c}(t)\right) + \mathbf{S}(\mathbf{u})\mathbf{r}(t;\mathbf{u}); \quad \left.\frac{d\mathbf{c}(t)}{dt}\right|_{t=t_0} = \mathbf{c}^I, \quad (S1)$$

where $\mathbf{c}(t) \in \mathbb{R}^{n_S}$ is a state vector of $n_S$ concentration variables $[\![mmol\ m^{-3}]\!]$ given by,

$$\mathbf{c}^T(t) = \begin{bmatrix} c_{DIC}\ c_{O_2}\ c_{NH_3}\ c_{C_L}\ c_{C_R}\ c_{N_R}\ c_{ⓈⓈ_{P\{1\}}}, \ldots, c_{ⓈⓈ_{P\{n_P\}}}\ c_{C_{P\{1\}}}, \ldots, c_{C_{P\{n_P\}}} \\ c_{ⓈⓈ_{B\{1\}}}, \ldots, c_{ⓈⓈ_{B\{n_B\}}}\ c_{ⓈⓈ_{C\{1\}}}, \ldots, c_{ⓈⓈ_{C\{n_C\}}} \end{bmatrix}, \quad (S2)$$

that consists of 6 chemical constituents, $n_P$ phytoplankton ecotypes and associated internal $C_P$ storage pool, $n_B$ bacteria ecotypes and $n_C$ consumer ecotypes so that $n_S = 6 + 2n_P + n_B + n_C$; $\mathbf{c}^I$ are the input concentrations that also serve as the initial conditions at $t_0$; a stagnant-film model governs mass exchange across the air-water interface for state variables with gas phases ($CO_2$ and $O_2$), where $\mathbf{v}$ $[\![m\ d^{-1}]\!]$ is the piston velocity, $\mathbf{p}$ $[\![Pa]\!]$ is atmospheric gas partial pressure and $\mathbf{h}(T)$ is the Henry's law coefficient $[\![mmol\ m^{-3}\ Pa^{-1}]\!]$ at temperature $T$ $[\![K]\!]$, and ∘ is the element-wise multiplication (Hadamard) operator; $\mathbf{r}(t;\mathbf{u}) \in \mathbb{R}^{n_r}$ is a vector of $n_r$ reaction rates $[\![mmol\ m^{-3}\ d^{-1}]\!]$ associated with biological structures (see below) and $\mathbf{S}(\mathbf{u}) \in \mathbb{R}^{n_s \times n_r}$ is a reaction stoichiometry matrix. Reaction rates and stoichiometric matrix depend on a time-invariant control vector, $\mathbf{u}$, that consists of a vector of reaction efficiencies, $\boldsymbol{\varepsilon}$, and a vector of resource allocation controls, $\boldsymbol{\Omega}$, ($\mathbf{u}^T = [\boldsymbol{\varepsilon}^T\ \boldsymbol{\Omega}^T]$) described in Section S2.2. In this formulation, the control variables, $\boldsymbol{\Omega}$ and $\boldsymbol{\varepsilon}$, are held constant for each ecotype, so serve as the trait variables.

*S2.2 Metabolic Reaction Rates*

The metabolic reactions associated with phytoplankton, bacteria and consumers (grazers) follows that developed previously [2], except in this implementation each functional group can have a specified number of ecotypes that have different values for the control variables (i.e., traits). We use braces, $\{i\}$, to designate each realization of an ecotype defined by the trait values and $\chi$ represents one of the three functional groups (P, B or C). The governing equation for each of the three functional groups are given below, with the following overall organization. The metabolic reactions a functional group is capable of catalyzing includes the thermodynamic efficiency trait, $\varepsilon_{\chi\{i\}}$, that specifies weighting between an anabolic (i.e., biosynthesis) reaction and a catabolic (energy producing) reaction. The anabolic and catabolic reactions are combined into a single reaction and balanced with the parameter $n_{j,\chi\{i\}}$ that ensures as $\varepsilon_{\chi\{i\}}$ approaches 1, the Gibbs free energy of reaction goes to 0. The anabolic and catabolic reactions can be recovered by setting $\varepsilon_{\chi\{i\}}$ to 1 or 0, respectively. Reaction entropy is maximized as $\varepsilon_{\chi\{i\}}$ approaches 0, as this represents complete destruction of the energy potential. Stoichiometric coefficients, such as $a_{2,P}^A$ and $b_{1,B}^C$, are used to balance O and H, respectively, where the superscript is for either the anabolic (A) or catabolic (C) reaction, and the subscripts correspond to the reaction number for the associated functional group (P, B or C). Defined by whole reaction stoichiometry, the Gibbs free energy of reaction, $\Delta_r G$, accounts for the reaction quotient, and the standard Gibbs free energy of reaction, $\Delta_r G^o$, is obtained from Alberty's [4] that accounts for ionization of chemical species based on pH, temperature and ionic strength to approximate activity from concentration. Reaction kinetics are based on an adaptive Monod equation [2] that is parameterized by $\varepsilon_{\chi\{i\}}$ and includes a thermodynamic driver, $F_T$, that depends on the number of electrons, $n_{j,\chi\{i\}}^e$, transferred in the catabolic reaction as described by LaRowe et al. [5], and bracket notation, [ ], is used to represent concentration of state variables (i.e., $[NH_3]$ and $c_{NH_3}$ are equivalent). The fraction of biological structure allocated to each metabolic reaction that a functional group can catalyzed is determined by $\Omega_{j,\chi\{i\}}$, where $\sum_j \Omega_{j,\chi\{i\}} = 1$ and $0 \leq \Omega_{j,\chi\{i\}} \leq 1\ \forall j$ because the total catalytic capacity is limited by the concertation of biological structure, $[\mathbb{S}_{\chi\{i\}}]$. Entropy production is calculated for dissipation of chemical potential by metabolic reactions, $\dot{\sigma}_{j,\chi\{i\}}^R$, as well as dissipation of electromagnetic potential by particulate material, $\dot{\sigma}_{j,\chi\{i\}}^P$, and water, $\dot{\sigma}^W$, although the latter does not depend on any of the state variables, so is not listed below (See Section S2.4 below).

S2.2.1 Phytoplankton Reactions

Phytoplankton are represented with two metabolic reactions consisting of 1) $CO_2$ fixation into unit-C sugar (i.e. $CH_2O$, or $C_{P\{i\}}$) driven by high frequency photon, $\gamma_H$, capture and 2) conversion of $C_{P\{i\}}$ into biomass using available ammonium and phosphate driven by the catabolic aerobic oxidation of $C_{P\{i\}}$. (Note, phosphate is not a state variable and is held at a fixed concentration of 1 μM during simulations.) Surficial light intensity, $I_0(t)$, varies on both diel and seasonal cycles [3], and depth-average light intensity, $\langle I(t) \rangle_d$, for the well-mixed system is calculated from $I_0(t)$ and light at depth $d$, where light attenuation occurs by water, particles and chlorophyll a as parameterized by $k_w$, $k_p$ and $k_{Chl}$ respectively. We only consider blue light at 440 nm and the light attenuation coefficients, $k_w$, $k_p$ and $k_{Chl}$ were derived from Wozniak [6] and set to 0.011 m$^{-1}$, 0.000625 m$^2$ (mmol-C)$^{-1}$ and 0.0025 m$^2$ (mmol-C)$^{-1}$ for 440 nm light, respectively, after conversion to mM C. The Gibbs free energy of photons, $\Delta_r G_\gamma$, at 440 nm is -253 J (mmol-$\gamma$)$^{-1}$, which accounts for the conversion of photons to work [2,7]. Entropy production for phytoplankton is divided into reaction associated, $\dot{\sigma}^R_{1,P\{i\}}$, and particle absorption, $\dot{\sigma}^P_{1,P\{i\}}$, components that are controlled by $\Omega_{1,P\{i\}}$ and $\Omega_{2,P\{i\}}$. Light absorbed by water and non-photosynthetic biomass is simply dissipated as heat and contributes to entropy production. Only the fraction of electromagnetic potential that is conserved as chemical potential (i.e., $C_{P\{i\}}$ and $\mathcal{S}_{P\{i\}}$ synthesis) does not contribute to entropy production (see Eqn. (S13) for $\dot{\sigma}^T_{1,P\{i\}}$ below). The fraction of phytoplankton biomass allocated to photosynthetic processes described by $r_{1,P\{i\}}$ is given by $\Omega_{1,P\{i\}}[\mathcal{S}_{P\{i\}}]$; however, since the photosynthetic machinery can be kinetically limited by resource availability (i.e., $[CO_2] + [HCO_3^-]$), only the fraction of the total photon capture rate that contributes to $r_{1,P\{i\}}$, given by $\frac{\Delta I_{P\{i\}}}{n_{1,P\{i\}}}$, contributes to $\dot{\sigma}^R_{1,P\{i\}}$, while the remainder contributes to $\dot{\sigma}^P_{1,P\{i\}}$. That is, if the photosynthetic machinery is constrained, the excess light captured is dissipated, so contributes to $\dot{\sigma}^P_{1,P\{i\}}$. Biological structure associated with metabolism and biosynthesis, given by $\Omega_{2,P\{i\}}[\mathcal{S}_{P\{i\}}]$ that controls $r_{2,P\{i\}}$, always contributes to particle-associated entropy production for light interception, while dissipated chemical potential associated with catabolic reactions contributes to reaction-associated entropy production, $\dot{\sigma}^R_{2,P\{i\}}$. For the carbon dioxide fixation reaction (Eq. (S3)), $n_{1,P\{i\}}$ is the moles of high frequency photons, $\gamma_H$, needed to fix one mole of $CO_2$, reversibly, under the current conditions, so that the quantum yield, $\frac{n_{1,P\{i\}}}{\varepsilon_{P\{i\}}}$, depends on $\varepsilon_{P\{i\}}$. The concentration of fixed carbon, $[C_{P\{i\}}]$, is based on total system volume, but for kinetics it is treated as an intracellular component, so is multiplied by a system-to-cell volume factor, $\varphi_f$, to reflect its higher intracellular concentration ($\varphi_f$ was set to 1000 for all simulations). Below are the equations describing phytoplankton growth and associated entropy production.

<u>Carbon dioxide fixation driven by solar radiation:</u> $r_{1,P\{i\}}$

$$\varepsilon_{P\{i\}} H_2CO_3 + n_{1,P\{i\}} \gamma_H \rightarrow \varepsilon_{P\{i\}}\left(C_{P\{i\}} + O_2(aq)\right) \tag{S3}$$

$$\Delta_r G^o_{C_{P\{i\}}} = \Delta_f G^o_{CH_2O} + \Delta_f G^o_{O_2(aq)} - \Delta_f G^o_{H_2CO_3} \tag{S4}$$

$$\Delta_r G_{C_{P\{i\}}} = \Delta_r G^o_{C_{P\{i\}}} + RT \ln\left(\frac{[C_{P\{i\}}][O_2(aq)]}{[H_2CO_3]}\right) \tag{S5}$$

$$n_{1,P\{i\}} = -\frac{\Delta_r G_{C_{P\{i\}}}}{\Delta_r G_\gamma} \tag{S6}$$

$$\Delta_r G_{1,P\{i\}} = -(1-\varepsilon_{P\{i\}})\Delta_r G_{C_{P\{i\}}} \tag{S7}$$

$$k_{wp} = k_w + k_p\left(\sum_i \Omega_{2,P\{i\}}[\mathcal{S}_{P\{i\}}] + \sum_i [C_{P\{i\}}] + \sum_i [\mathcal{S}_{B\{i\}}] + \sum_i [\mathcal{S}_{C\{i\}}]\right) + k_{Chl} \sum_i \Omega_{1,P\{i\}}[\mathcal{S}_{P\{i\}}] \tag{S8}$$

$$\langle I(t)\rangle_d = \frac{I_0(t)(1-e^{-k_{wp}d})}{k_{wp}d} \tag{S9}$$

$$\Delta I_{P\{i\}} = k_{Chl} \Omega_{1,P\{i\}}[\mathcal{S}_{P\{i\}}]\langle I(t)\rangle_d \tag{S10}$$

$$n^e_{1,P\{i\}} = n_{1,P\{i\}} \tag{S11}$$

$$r_{1,P\{i\}} = \frac{\Delta I_{P\{i\}}}{n_{1,P\{i\}}}\left(\frac{[CO_2] + [HCO_3^-]}{[CO_2] + [HCO_3^-] + \kappa^* \varepsilon^4_{P\{i\}}}\right) F_T(\Delta_r G_{1,P\{i\}}, n^e_{1,P\{i\}}) \tag{S12}$$

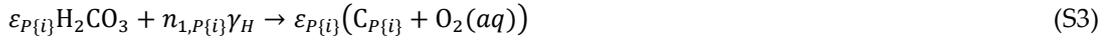

$$\dot{\sigma}_{1,P\{i\}}^T = \frac{Ad}{T}\Delta_r G_{C_{P\{i\}}}\left(\frac{\Delta I_{P\{i\}}}{n_{1,P\{i\}}} - \varepsilon_{P\{i\}} r_{1,P\{i\}}\right) \tag{S13}$$

$$\dot{\sigma}_{1,P\{i\}}^R = \frac{r_{1,P\{i\}} n_{1,P\{i\}}}{\Delta I_{P\{i\}}}\dot{\sigma}_{1,P\{i\}}^T \tag{S14}$$

$$\dot{\sigma}_{1,P\{i\}}^P = \dot{\sigma}_{1,P\{i\}}^T - \dot{\sigma}_{1,P\{i\}}^R \tag{S15}$$

Conversion of reduced carbon into phytoplankton biomass: $r_{2,P\{i\}}$

$$\begin{aligned}(1 + \varepsilon_{P\{i\}} n_{2,P\{i\}})C_{P\{i\}} + \varepsilon_{P\{i\}}(\gamma_{\text{\$}}NH_3 + \delta_{\text{\$}}H_3PO_4) + \left(1 + \varepsilon_{P\{i\}}(a_{2,P}^A + n_{2,P\{i\}} - 1)\right)O_2(aq) \\ \to \varepsilon_{P\{i\}}\text{\$}_{P\{i\}} + \varepsilon_{P\{i\}} b_{2,P}^A H_2O + \left(1 + \varepsilon_{P\{i\}}(n_{2,P\{i\}} - 1)\right)H_2CO_3\end{aligned} \tag{S16}$$

$$a_{2,P}^A = \tfrac{1}{4}(-\alpha_{\text{\$}} + 2\beta_{\text{\$}} + 3\gamma_{\text{\$}} - 5\delta_{\text{\$}}) \tag{S17}$$

$$b_{2,P}^A = \tfrac{1}{2}(2 - \alpha_{\text{\$}} + 3\gamma_{\text{\$}} + 3\delta_{\text{\$}}) \tag{S18}$$

$$\begin{aligned}\Delta_r^A G_{2,P\{i\}}^o &= \left(\Delta_f G_{\text{\$}}^o + b_{2,P}^A \Delta_f G_{H_2O}^o\right) \\ &- \left(\Delta_f G_{C_{P\{i\}}}^o + \gamma_{\text{\$}}\Delta_f G_{NH_3}^o + \delta_{\text{\$}}\Delta_f G_{H_3PO_4}^o + a_{2,P}^A \Delta_f G_{O_2(aq)}^o\right)\end{aligned} \tag{S19}$$

$$\Delta_r^A G_{2,P\{i\}} = \Delta_r^A G_{2,P\{i\}}^o + RT\ln\left(\frac{[\text{\$}_{P\{i\}}]}{[C_{Phy}][NH_3]^{\gamma_{\text{\$}}}[H_3PO_4]^{\delta_{\text{\$}}}[O_2(aq)]^{a_{2,P}^A}}\right) \tag{S20}$$

$$\Delta_r^C G_{2,P\{i\}}^o = \Delta_f G_{H_2CO_3}^o - \left(\Delta_f G_{C_{P\{i\}}}^o + \Delta_f G_{O_2(aq)}^o\right) \tag{S21}$$

$$\Delta_r^C G_{2,P\{i\}} = \Delta_r^C G_{2,P\{i\}}^o + RT\ln\left(\frac{[H_2CO_3]}{[C_{P\{i\}}][O_2(aq)]}\right) \tag{S22}$$

$$n_{2,P\{i\}} = -\frac{\Delta_r^A G_{2,P\{i\}}}{\Delta_r^C G_{2,P\{i\}}} \tag{S23}$$

$$\Delta_r G_{2,P\{i\}} = \left(1 - \varepsilon_{P\{i\}}\right)\Delta_r^C G_{2,P\{i\}}$$

$$n_{2,P\{i\}}^e = 4 \tag{S24}$$

$$\begin{aligned}r_{2,P\{i\}} = \nu^* \varepsilon_{P\{i\}}^2 \Omega_{2,P\{i\}}[\text{\$}_{P\{i\}}]\left(\frac{[C_{P\{i\}}]\varphi_f}{[C_{P\{i\}}]\varphi_f + \kappa^*\varepsilon_{P\{i\}}^4}\right)\left(\frac{[NH_3]/\gamma_{\text{\$}}}{[NH_3]/\gamma_{\text{\$}} + \kappa^*\varepsilon_{P\{i\}}^4}\right) \\ \times \left(\frac{[O_2]}{[O_2] + \kappa^*\varepsilon_{P\{i\}}^4}\right)F_T\left(\Delta_r G_{2,P\{i\}}, n_{2,P\{i\}}^e\right)\end{aligned} \tag{S25}$$

$$\dot{\sigma}_{2,P\{i\}}^R = -\frac{Ad}{T}r_{2,P\{i\}}\Delta_r G_{2,P\{i\}} \tag{S26}$$

$$\dot{\sigma}_{2,P\{i\}}^P = -\frac{Ad}{T}\langle I(t)\rangle_d \Delta_r G_\gamma k_p \Omega_{2,P\{i\}}[\text{\$}_{P\{i\}}] \tag{S27}$$

S2.2.2 Bacteria growth on labile carbon, $C_L$

Bacteria catalyze three reactions that include growth on labile carbon, $C_L$, and decomposition of detrital carbon, $C_D$, and nitrogen, $N_D$, into labile pools, where $\Omega_{1,B\{i\}}$, $\Omega_{2,B\{i\}}$ and $\Omega_{3,B\{i\}}$ determine the allocation of catalytic machinery to each reaction, respectively. Decomposition of detritus, which is recalcitrant, uses a different biomass-specific rate constant, $v_D^*$, than that used for growth on labile carbon.

Bacterial growth: $r_{1,B\{i\}}$

$$C_L + \varepsilon_{B\{i\}}(\gamma_{\circleddash} NH_3 + \delta_{\circleddash} H_3PO_4) + (1 - \varepsilon_{B\{i\}})O_2(aq)$$
$$\rightarrow \varepsilon_{B\{i\}} a_{1,B}^A \circleddash_{B\{i\}} + \left(2 - \varepsilon_{B\{i\}}(a_{1,B}^A + 1)\right) H_2CO_3 + \varepsilon_{B\{i\}} b_{1,B}^A H_2O \quad \text{(S28)}$$

$$a_{1,B}^A = \frac{4 + 3\gamma_{\circleddash} - 5\delta_{\circleddash}}{4 + \alpha_{\circleddash} - 2\beta_{\circleddash}} \quad \text{(S29)}$$

$$b_{1,B}^A = \frac{4 - 2\alpha_{\circleddash} + 9\gamma_{\circleddash} - 3\beta_{\circleddash}\gamma_{\circleddash} + \delta_{\circleddash} + 4\alpha_{\circleddash}\delta_{\circleddash} - 3\beta_{\circleddash}\delta_{\circleddash}}{4 + \alpha_{\circleddash} - 2\beta_{\circleddash}} \quad \text{(S30)}$$

$$\Delta_r^A G_{1,B\{i\}}^o = \left(a_{1,B}^A \Delta_f G_{\circleddash}^o + (1 - a_{1,B}^A)\Delta_f G_{H_2CO_3}^o + b_{1,B}^A \Delta_f G_{H_2O}^o\right)$$
$$- \left(\Delta_f G_{C_L}^o + \gamma_{\circleddash}\Delta_f G_{NH_3}^o + \delta_{\circleddash}\Delta_f G_{H_3PO_4}^o\right) \quad \text{(S31)}$$

$$\Delta_r^A G_{1,B\{i\}} = \Delta_r^A G_{1,B\{i\}}^o + RT\ln\left(\frac{[\circleddash_{B\{i\}}]^{a_{1,B}^A}[H_2CO_3]^{(1-a_{1,B}^A)}}{[C_L][NH_3]^{\gamma_{\circleddash}}[H_3PO_4]^{\delta_{\circleddash}}}\right) \quad \text{(S32)}$$

$$\Delta_r^C G_{1,B\{i\}}^o = \Delta_f G_{H_2CO_3}^o - \left(\Delta_f G_{C_L}^o + \Delta_f G_{O_2(aq)}^o\right) \quad \text{(S33)}$$

$$\Delta_r^C G_{1,B\{i\}} = \Delta_r^C G_{1,B\{i\}}^o + RT\ln\left(\frac{[H_2CO_3]}{[C_L][O_2(aq)]}\right) \quad \text{(S34)}$$

$$\Delta_r G_{1,B\{i\}} = \varepsilon_{B\{i\}}\Delta_r^A G_{1,B\{i\}} + (1 - \varepsilon_{B\{i\}})\Delta_r^C G_{1,B\{i\}} \quad \text{(S35)}$$

$$n_{1,B\{i\}}^e = 4 \quad \text{(S36)}$$

$$r_{1,B\{i\}} = v^* \varepsilon_{B\{i\}}^2 \Omega_{1,B\{i\}}[\circleddash_{B\{i\}}]\left(\frac{[C_L]}{[C_L] + \kappa^* \varepsilon_{B\{i\}}^4}\right)\left(\frac{[NH_3]/\gamma_{\circleddash}}{[NH_3]/\gamma_{\circleddash} + \kappa^* \varepsilon_{B\{i\}}^4}\right)$$
$$\times \left(\frac{[O_2]}{[O_2] + \kappa^* \varepsilon_{B\{i\}}^4}\right) F_T(\Delta_r G_{1,B\{i\}}, n_{1,B\{i\}}^e) \quad \text{(S37)}$$

$$\dot{\sigma}_{1,B\{i\}}^R = -\frac{Ad}{T} r_{1,B\{i\}} \Delta_r G_{1,B\{i\}} \quad \text{(S38)}$$

$$\dot{\sigma}_{1,B\{i\}}^P = -\frac{Ad}{T} \langle I(t) \rangle_d \Delta_r G_\gamma k_p \Omega_{1,B\{i\}}[\circleddash_{B\{i\}}] \quad \text{(S39)}$$

Bacterial decomposition of recalcitrant carbon: $r_{2,B\{i\}}$

$$C_D \rightarrow C_L \quad \text{(S40)}$$

$$\Delta_r G_{2,B\{i\}} = RT\ln\left(\frac{[C_L]}{[C_D]}\right) \quad \text{(S41)}$$

$$r_{2,B\{i\}} = v_D^* \varepsilon_{B\{i\}}^2 \Omega_{2,B\{i\}}[\circleddash_{B\{i\}}]\left(\frac{[C_D]}{[C_D] + \kappa^* \varepsilon_{B\{i\}}^4}\right), \text{for } \Delta_r G_{2,B\{i\}} < 0; 0 \text{ otherise} \quad \text{(S42)}$$

$$\dot{\sigma}_{2,B\{i\}}^R = -\frac{Ad}{T} r_{2,B\{i\}} \Delta_r G_{2,B\{i\}} \quad \text{(S43)}$$

$$\dot{\sigma}_{2,B\{i\}}^P = -\frac{Ad}{T}\langle I(t)\rangle_d \Delta_r G_\gamma k_p \Omega_{2,B\{i\}}[\text{\S}_{B\{i\}}] \tag{S44}$$

Bacterial decomposition of recalcitrant nitrogen: $r_{3,B\{i\}}$

$$N_D \to NH_3 \tag{S45}$$

$$\Delta_r G_{3,B\{i\}} = RT \ln\left(\frac{[NH_3]}{[N_D]}\right) \tag{S46}$$

$$r_{3,B\{i\}} = v_D^* \varepsilon_{B\{i\}}^2 \Omega_{3,B\{i\}}[\text{\S}_{B\{i\}}]\left(\frac{[N_D]}{[N_D] + \kappa^* \varepsilon_{Bac}^4}\right), \text{for } \Delta_r G_{3,B\{i\}} < 0; 0 \text{ otherise} \tag{S47}$$

$$\dot{\sigma}_{3,B\{i\}}^R = -\frac{Ad}{T} r_{3,B\{i\}} \Delta_r G_{3,B\{i\}} \tag{S48}$$

$$\dot{\sigma}_{3,B\{i\}}^P = -\frac{Ad}{T}\langle I(t)\rangle_d \Delta_r G_\gamma k_p \Omega_{3,B\{i\}}[\text{\S}_{B\{i\}}] \tag{S49}$$

S2.2.3 Consumer predation rate, $r_{\chi\{j\},C\{i\}}$

Consumers prey on all function groups including themselves (i.e., cannibalism). We use nearly the same governing equations as before [2] where C in biological structure is converted into biomass, DIC and detrital carbon; however, all $C_{P\{i\}}$ storage in phytoplankton is oxidized to $H_2CO_3$ instead of being excreted as $C_L$. Excess N and P from consumed biological structure is excreted in both labile and detrital forms as a function of $\varepsilon_{C\{i\}}$. The rational is that when a consumer operates with high thermodynamic efficiency ($\varepsilon_{C\{i\}}$ closer to 1), prey are processed more effectively leading to $NH_3$ and $H_3PO_4$ production, while low efficiency growth leads to more detrital products. This version also weights allocation to prey consumption normalized by prey density, as given by $\omega_{\chi\{j\},C\{i\}}$ below; consequently, no constraint is placed on the sum, $\sum_j \Omega_{\chi\{j\},C\{i\}}$, as it is for phytoplankton and bacteria. In the equations below, the subscript $\chi\{j\}$ represented any instance, $\{j\}$, of any of the three functional groups, where $\chi$ can be P, B or C, and $[C_{\chi\{j\}}]$ equals 0 for $\chi$ equal to B or C, since those functional groups have no internal carbon storage.

$$\begin{aligned}
\text{\S}_{\chi\{j\}} + \frac{[C_{\chi\{j\}}]}{[\text{\S}_{\chi\{j\}}]} C_{\chi\{j\}} &+ \left(a_{C\{i\}}^C(1-\varepsilon_{C\{i\}}) + \frac{[C_{\chi\{j\}}]}{[\text{\S}_{\chi\{j\}}]}\right) O_2(aq) \\
\to \varepsilon_{C\{i\}} &\text{\S}_{C\{i\}} + (1-\varepsilon_{C\{i\}})\left((1-\varepsilon_{C\{i\}})H_2CO_3 + \varepsilon_{C\{i\}}C_D\right) \\
&+ \gamma_\text{\S}(1-\varepsilon_{C\{i\}})\left((1-\varepsilon_{C\{i\}})NH_3 + \varepsilon_{C\{i\}}N_D\right) \\
&+ \delta_\text{\S}(1-\varepsilon_{C\{i\}})\left((1-\varepsilon_{C\{i\}})H_3PO_4 + \varepsilon_{C\{i\}}P_D\right) + b_C^C(1-\varepsilon_{C\{i\}})H_2O \\
&+ \frac{[C_{\chi\{j\}}]}{[\text{\S}_{\chi\{j\}}]} H_2CO_3
\end{aligned} \tag{S50}$$

$$a_{C\{i\}}^C = \tfrac{1}{4}(4 + \alpha_\text{\S} - 2\beta_\text{\S} - 3\gamma_\text{\S} + 5\delta_\text{\S} - 4\varepsilon_{C\{i\}}) \tag{S51}$$

$$b_C^C = \tfrac{1}{2}(-2 + \alpha_\text{\S} - 3\gamma_\text{\S} - 3\delta_\text{\S}) \tag{S52}$$

$$\Delta_r G^o_{C\{i\}} = \left( \varepsilon_{C\{i\}} \Delta_f G^o_{\maltese} + \left( (1-\varepsilon_{C\{i\}})^2 + \frac{[C_{\chi\{j\}}]}{[\maltese_{\chi\{j\}}]} \right) \Delta_f G^o_{H_2CO_3} + \varepsilon_{C\{i\}}(1-\varepsilon_{C\{i\}}) \Delta_f G^o_{C_D} \right.$$
$$+ \delta_{\maltese}(1-\varepsilon_{C\{i\}}) \Delta_f G^o_{H_3PO_4} + \gamma_{\maltese}(1-\varepsilon_{C\{i\}}) \Delta_f G^o_{NH_3}$$
$$\left. + b^C_C(1-\varepsilon_{C\{i\}}) \Delta_f G^o_{H_2O} \right)$$
$$- \left( \Delta_f G^o_{\maltese} + \left( a^C_{C\{i\}}(1-\varepsilon_{C\{i\}}) + \frac{[C_{\chi\{j\}}]}{[\maltese_{\chi\{j\}}]} \right) \Delta_f G^o_{O_2(aq)} + \frac{[C_{\chi\{j\}}]}{[\maltese_{\chi\{j\}}]} \Delta_f G^o_{CH_2O} \right)$$
(S53)

$$\Delta_r G_{\chi\{j\},C\{i\}} = \Delta_r G^o_{C\{i\}} + RTln\left( [\maltese_{C\{i\}}]^{\varepsilon_{C\{i\}}} [H_2CO_3]^{(1-\varepsilon_{C\{i\}})^2} [C_D]^{\varepsilon_{C\{i\}}(1-\varepsilon_{C\{i\}})} \right)$$
$$+ RTln\left( [NH_3]^{\gamma_{\maltese}(1-\varepsilon_{C\{i\}})^2} [N_D]^{\gamma_{\maltese}\varepsilon_{C\{i\}}(1-\varepsilon_{C\{i\}})} \right)$$
$$+ RTln\left( [H_3PO_4]^{\delta_{\maltese}(1-\varepsilon_{C\{i\}})^2} [P_D]^{\delta_{\maltese}\varepsilon_{C\{i\}}(1-\varepsilon_{C\{i\}})} \right)$$
$$- RTln\left( [\maltese_{\chi\{j\}}][C_{\chi\{j\}}]^{[C_{\chi\{j\}}]/[\chi\{j\}]} [O_2(aq)]^{a^C_{C\{i\}}(1-\varepsilon_{C\{i\}})} \right)$$
(S54)

$$n^e_{C\{i\}} = 4 \tag{S55}$$

$$\omega_{\chi\{j\},C\{i\}} = \frac{\Omega_{\chi\{j\},C\{i\}}[\maltese_{\chi\{j\}}]}{\sum_j \Omega_{P\{j\},C\{i\}}[\maltese_{P\{j\}}] + \sum_j \Omega_{B\{j\},C\{i\}}[\maltese_{B\{j\}}] + \sum_j \Omega_{C\{j\},C\{i\}}[\maltese_{C\{j\}}]} \tag{S56}$$

$$r_{\chi\{j\},C\{i\}}$$
$$= \nu^* \varepsilon^2_{C\{i\}} \omega_{\chi\{j\},C\{i\}}[\maltese_{C\{i\}}] \left( \frac{[\maltese_{\chi\{j\}}]}{[\maltese_{\chi\{j\}}] + \kappa^* \varepsilon^4_{C\{i\}}} \right) \left( \frac{[O_2(aq)]}{[O_2(aq)] + \kappa^* \varepsilon^4_{C\{i\}}} \right) F_T(\Delta_r G_{\chi\{j\},C\{i\}}, n^e_{C\{i\}}) \tag{S57}$$

$$\dot{\sigma}^R_{\chi\{j\},C\{i\}} = -\frac{Ad}{T} r_{\chi\{j\},C\{i\}} \Delta_r G_{\chi\{j\},C\{i\}} \tag{S58}$$

$$\dot{\sigma}^P_{\chi\{j\},C\{i\}} = -\frac{Ad}{T} \langle I(t) \rangle_d \Delta_r G_\gamma k_p \Omega_{\chi\{j\},C\{i\}}[\maltese_{C\{i\}}] \tag{S59}$$

*S2.3 Reaction Network, $\mathbf{S(u)r}(t;\mathbf{u})$*

The reaction network is defined by the stoichiometries of the reactions listed above. Instead of listing all the elements of the stoichiometric matrix, $\mathbf{S(u)}$, which is sparse, we list the $n_S$ rows of the vector that results from the matrix vector product of $\mathbf{S(u)r}(t;\mathbf{u})$, which is a mass balance around each state variable. For instance, $\mathbf{S}^T_{\maltese_{P\{i\}}}\mathbf{r}$ is the net production rate of $\maltese_{P\{j\}}$ resulting from the growth and predation.

$$\mathbf{S}^T_{DIC}\mathbf{r} = \sum_{i=1}^{n_P} \left( -\varepsilon_{P\{i\}} r_{1,P\{i\}} + \left(1 + \varepsilon_{P\{i\}}(n_{2,P\{i\}} - 1)\right) r_{2,P\{i\}} \right)$$
$$+ \sum_{i=1}^{n_B} \left( 2 - \varepsilon_{B\{i\}}(a^A_{1,B} + 1) \right) r_{1,B\{i\}}$$
$$+ \sum_{i=1}^{n_C} \sum_{j=1}^{n_P} \left( (1-\varepsilon_{C\{i\}})^2 + \frac{[C_{P\{j\}}]}{[\maltese_{P\{j\}}]} \right) r_{P\{j\},C\{i\}} + \sum_{i=1}^{n_C} \sum_{j=1}^{n_B} (1-\varepsilon_{C\{i\}})^2 r_{B\{j\},C\{i\}}$$
$$+ \sum_{i=1}^{n_C} \sum_{j=1}^{n_C} (1-\varepsilon_{C\{i\}})^2 r_{C\{j\},C\{i\}}$$
(S60)

$$\mathbf{S}_{O_2}^T \mathbf{r} = \sum_{i=1}^{n_P} \left( \varepsilon_{P\{i\}} r_{1,P\{i\}} - \left(1 + \varepsilon_{P\{i\}} \left(a_{2,P}^A + n_{2,P\{i\}} - 1\right)\right) r_{2,P\{i\}} \right) - \sum_{i=1}^{n_B} (1 - \varepsilon_{B\{i\}}) r_{1,B\{i\}}$$
$$- \sum_{i=1}^{n_C} \sum_{j=1}^{n_P} \left( a_{C\{i\}}^C (1 - \varepsilon_{C\{i\}}) + \frac{[C_{P\{j\}}]}{[\mathcal{S}_{P\{j\}}]} \right) r_{P\{j\},C\{i\}} \quad (S61)$$
$$- \sum_{i=1}^{n_C} \sum_{j=1}^{n_B} a_{C\{i\}}^C (1 - \varepsilon_{C\{i\}}) r_{B\{j\},C\{i\}} - \sum_{i=1}^{n_C} \sum_{j=1}^{n_C} a_{C\{i\}}^C (1 - \varepsilon_{C\{i\}}) r_{C\{j\},C\{i\}}$$

$$\mathbf{S}_{C_L}^T \mathbf{r} = \sum_{i=1}^{n_B} \left( -r_{1,B\{i\}} + r_{2,B\{i\}} \right) \quad (S62)$$

$$\mathbf{S}_{C_D}^T \mathbf{r} = -\sum_{i=1}^{n_B} r_{2,B\{i\}} + \sum_{i=1}^{n_C} \sum_{j=1}^{n_P} \varepsilon_{C\{i\}} (1 - \varepsilon_{C\{i\}}) r_{P\{j\},C\{i\}} + \sum_{i=1}^{n_C} \sum_{j=1}^{n_B} \varepsilon_{C\{i\}} (1 - \varepsilon_{C\{i\}}) r_{B\{j\},C\{i\}}$$
$$+ \sum_{i=1}^{n_C} \sum_{j=1}^{n_C} \varepsilon_{C\{i\}} (1 - \varepsilon_{C\{i\}}) r_{C\{j\},C\{i\}} \quad (S63)$$

$$\mathbf{S}_{N_D}^T \mathbf{r} = -\sum_{i=1}^{n_B} r_{3,B\{i\}} + \sum_{i=1}^{n_C} \sum_{j=1}^{n_P} \gamma_{\mathcal{S}} \varepsilon_{C\{i\}} (1 - \varepsilon_{C\{i\}}) r_{P\{j\},C\{i\}}$$
$$+ \sum_{i=1}^{n_C} \sum_{j=1}^{n_B} \gamma_{\mathcal{S}} \varepsilon_{C\{i\}} (1 - \varepsilon_{C\{i\}}) r_{B\{j\},C\{i\}} + \sum_{i=1}^{n_C} \sum_{j=1}^{n_C} \gamma_{\mathcal{S}} \varepsilon_{C\{i\}} (1 - \varepsilon_{C\{i\}}) r_{C\{j\},C\{i\}} \quad (S63)$$

$$\mathbf{S}_{NH_3}^T \mathbf{r} = -\sum_{i=1}^{n_P} \varepsilon_{P\{i\}} \gamma_{\mathcal{S}} r_{2,P\{i\}} + \sum_{i=1}^{n_B} \left( r_{3,B\{i\}} - \varepsilon_{B\{i\}} \gamma_{\mathcal{S}} r_{1,B\{i\}} \right) + \sum_{i=1}^{n_C} \sum_{j=1}^{n_P} \gamma_{\mathcal{S}} (1 - \varepsilon_{C\{i\}})^2 r_{P\{j\},C\{i\}}$$
$$+ \sum_{i=1}^{n_C} \sum_{j=1}^{n_B} \gamma_{\mathcal{S}} (1 - \varepsilon_{C\{i\}})^2 r_{B\{j\},C\{i\}} + \sum_{i=1}^{n_C} \sum_{j=1}^{n_C} \gamma_{\mathcal{S}} (1 - \varepsilon_{C\{i\}})^2 r_{C\{j\},C\{i\}} \quad (S65)$$

$$\mathbf{S}_{\mathcal{S}_{P\{i\}}}^T \mathbf{r} = \varepsilon_{P\{i\}} r_{2,P\{i\}} - \sum_{j=1}^{n_C} r_{P\{i\},C\{j\}} \quad (S66)$$

$$\mathbf{S}_{C_{P\{i\}}}^T \mathbf{r} = \varepsilon_{P\{i\}} r_{1,P\{i\}} - \left(1 + \varepsilon_{P\{i\}} n_{2,P\{i\}}\right) r_{2,P\{i\}} - \sum_{j=1}^{n_C} \frac{[C_{P\{i\}}]}{[\mathcal{S}_{P\{i\}}]} r_{P\{i\},C\{j\}} \quad (S67)$$

$$\mathbf{S}_{\mathcal{S}_{B\{i\}}}^T \mathbf{r} = \varepsilon_{B\{i\}} a_{1,B}^A r_{1,B\{i\}} - \sum_{j=1}^{n_C} r_{B\{i\},C\{j\}} \quad (S68)$$

$$\mathbf{S}_{\mathcal{S}_{C\{i\}}}^T \mathbf{r} = \sum_{j=1}^{n_P} \varepsilon_{C\{i\}} r_{P\{j\},C\{i\}} + \sum_{j=1}^{n_B} \varepsilon_{C\{i\}} r_{B\{j\},C\{i\}} + \sum_{j=1}^{n_C} \varepsilon_{C\{i\}} r_{C\{j\},C\{i\}} - \sum_{j=1}^{n_C} r_{C\{i\},C\{j\}} \quad (S69)$$

*S2.4 Integrated Entropy Production*

Cumulative entropy production over the simulation (or optimization) interval is determined by summing then integrating the contributions of reactions, particle absorptions and water, as given by,

$$\sigma^R = \int_{t_0}^{t_f} \sum_{i=1}^{n_P} \left(\dot\sigma^R_{1,P\{i\}}(\tau) + \dot\sigma^R_{2,P\{i\}}(\tau)\right) d\tau + \int_{t_0}^{t_f} \sum_{i=1}^{n_B} \left(\dot\sigma^R_{1,B\{i\}}(\tau) + \dot\sigma^R_{2,B\{i\}}(\tau) + \dot\sigma^R_{3,B\{i\}}(\tau)\right) d\tau \\ + \int_{t_0}^{t_f} \sum_{j=1}^{n_\chi} \sum_{i=1}^{n_C} \dot\sigma^R_{\chi\{j\},C\{i\}}(\tau) \, d\tau \quad (S70)$$

$$\sigma^P = \int_{t_0}^{t_f} \sum_{i=1}^{n_P} \left(\dot\sigma^P_{1,P\{i\}}(\tau) + \dot\sigma^P_{2,P\{i\}}(\tau)\right) d\tau + \int_{t_0}^{t_f} \sum_{i=1}^{n_B} \left(\dot\sigma^P_{1,B\{i\}}(\tau) + \dot\sigma^P_{2,B\{i\}}(\tau) + \dot\sigma^P_{3,B\{i\}}(\tau)\right) d\tau \\ + \int_{t_0}^{t_f} \sum_{j=1}^{n_\chi} \sum_{i=1}^{n_C} \dot\sigma^P_{\chi\{j\},C\{i\}}(\tau) d\tau \quad (S71)$$

$$\sigma^W = \int_{t_0}^{t_f} \dot\sigma^W(\tau) d\tau \quad (S72)$$

The total entropy production used in the optimization described below is simply,

$$\sigma^T = \sigma^R + \sigma^W + \sigma^P \quad (S73)$$

*S2.5 Initial Value Problem Integration*

The numerical package BiM [8], which uses blended implicit methods to integrate stiff ordinary differential equations (ODEs), was used to solve the initial value problem (Eqn. S1) and to determine cumulative entropy production, Eqn. (S73). All simulations were run for two years, $10^{-6}$ was used for both absolute and relative tolerances, a maximum step size (*hmax*) of 0.05 was implemented to insure diel light cycles were not stepped over, *maxstep* was increased to 10000000, and finite differences were used to calculate the Jacobian matrix. Default values were used for all other BiM options.

**S3 Optimization of Trait-Based Model**

Instead of employing optimal control to determine how $\varepsilon_{\chi\{i\}}$ and $\Omega_{j,\chi\{i\}}$ vary over time as has been previously used [2,9], in this manuscript we investigated a hybrid between MEP optimization and trait-based modeling. In typical trait-based models [1], a large number of each functional group are included in the model, and the traits, (i.e., $\varepsilon_{\chi\{i\}}$ and $\Omega_{j,\chi\{i\}}$) are randomly assigned values. When a simulation is run, organisms that grow fastest under the prevailing simulated environment dominate, while others are effectively culled from the population in a manner analogous to natural selection, but *in silico*. In order to explore the trait space, a large population of each functional group is needed; however, this presents a problem in our current model formulation. As the population size of $\mathfrak{S}_P$ and $\mathfrak{S}_B$ are increased, the column dimension of the predation matrix, $\Omega_{\chi\{j\},C\{i\}}$, increases, so that the total number of traits in the model, given by $n_T = 2n_P + 3n_B + (1 + n_P + n_B + n_C)n_C$, increases rapidly with population size. For instance, a model with just 10 ecotypes in each functional group has 360 trait values in total, and one with 100 ecotypes each has 30,600 trait values in total; the size of the trait space scales with $O(n^2)$. One way to circumvent the scaling problem is to limit the number of prey each consumer can target, but this places more constraints on the structure of the food web than we desired. While we investigated the standard trait-based approach, we found the $O(n^2)$ scaling on trait space made the approach untenable for our objectives; consequently, we developed a new, hybrid approach.

Instead of randomly assigning trait values, the hybrid approach numerically searches for trait values that maximize an objective function, in this case entropy production. This approach does not require a large number of ecotypes of each functional group, because the trait space is not being explored randomly, but systematically using optimization. Even simulations with just one instance of each functional group ($n_T = 9$) generated reasonable solutions. In fact, as discussed in the main text, adding food web complexity in the form of more ecotypes did significantly increase EP in many of our 0D simulations. The hybrid approach differs from the optimal control approach in that neither $\varepsilon_{\chi\{i\}}$ nor $\Omega_{j,\chi\{i\}}$ vary during a simulation. Consequently, the size/complexity of the food web likely needs to be larger for temporally and spatially varying environments, but this was not invested in this study. Once optimal parameter values are determined, the optimization component does not need to be rerun.

### S3.1 hyperBOB

For the optimization, we used the derivative-free, box-constrained, local optimizer BOBYQA[10] to search the $n_T$ dimensional trait space by maximizing $\sigma^T$ defined by Eqns. (S70-S73). To search for a global optimum on a computer cluster with $N_{CPU}$ CPUs, BOBYQA was started with $N_{CPU}$ different initial conditions that were selected by sampling from a Latin unit hypercube [11], which we implemented in the routine hyperBOB (DOI: 10.5281/zenodo.3978689). Parameters used in BOBYQA/hyperBOB were: *rhobeg*, 0.49; *rhoend*, 0.0001; *maxfun*, 10,000. Except on rare occasions, solutions were found before the maximum number of function calls (*maxfun*) occurred. A time constraint was also placed on the solution, but it also seldom was invoked. All simulations were run on a 5-node computer cluster with 90 CPU cores.

### S3.2 Temporal Strategies for Phytoplankton

To investigate how temporal strategies, in particular circadian rhythms, increase entropy production, we used two different approaches. In the first approach that was later retired, the constant trait value assigned to $\Omega_{1,P\{i\}}$ was replaced by a time varying function that depends to two new trait variables, $f_{P\{i\}}$ and $\varphi_{P\{i\}}$, as given by,

$$\Omega_{1,P\{i\}}(t) = \frac{1}{2}\big(\sin(2\pi f_{P\{i\}} t + \varphi_{P\{i\}}) + 1\big), \tag{S74}$$

where $f_{P\{i\}}$ is the frequency $[\![d^{-1}]\!]$ and $\varphi_{P\{i\}}$ the phase $[\![rad]\!]$ of $\Omega_{1,P\{i\}}(t)$ that controls allocation of phytoplankton protein to CO$_2$ fixation given by reaction $r_{1,P\{i\}}$. When $f_{P\{i\}} = 0$, $\varphi_{P\{i\}}$ modifies the amplitude of $\Omega_{1,P\{i\}}$, but does not change over time. These two traits had the following bounds,

$$0 \leq f_{P\{i\}} \leq 2 \text{ and } 0 \leq \varphi_{P\{i\}} \leq 2\pi. \tag{S75}$$

Several simulations studies were conducted with the above temporal modification of $\Omega_{1,P\{i\}}(t)$; however, by allowing the frequency parameter to be a trait variable, we found that locating the global optimum proved challenging as evident in Figure S1, in which all parameters where held constant for a $1P1B1C$ simulation and $\sigma^R$ was calculated for different values of $f_{P\{1\}}$ and $\varphi_{P\{1\}}$ at high resolution (1051 uniform samples in each dimension). Even though local optima occur for other frequencies, all simulations investigated showed the global optimum only occurred for $f_{P\{1\}} = 1 \, d^{-1}$; consequently, we used a different function for $\Omega_{1,P\{i\}}(t)$ in which frequency was fixed to the diel cycle of 1 per day to improve computational speed.

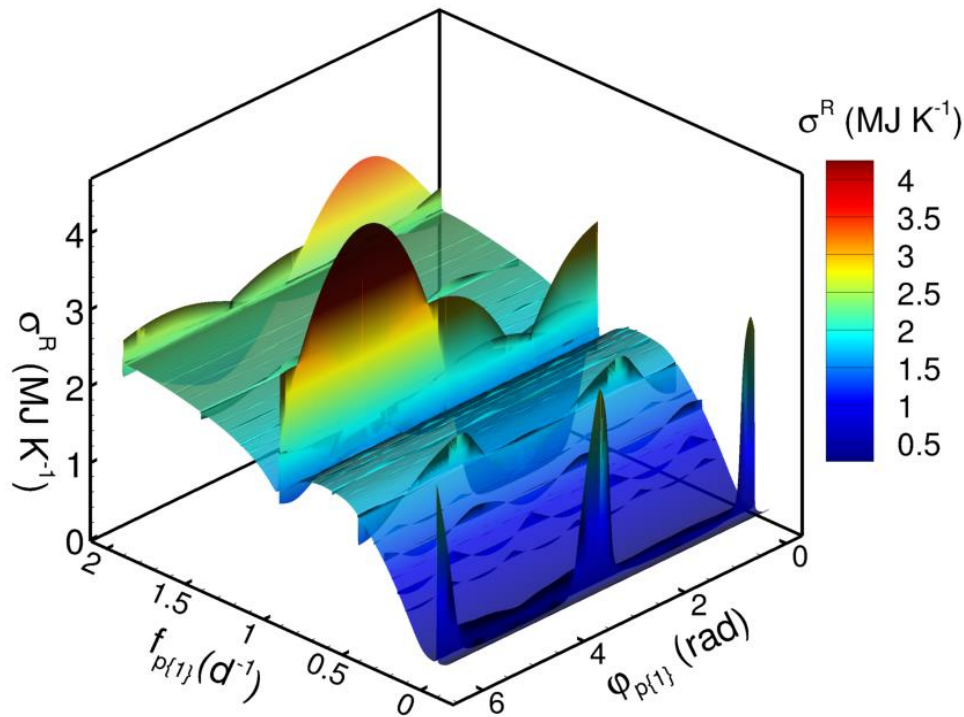

Figure S1. Reaction entropy production, $\sigma^R$, for different values of $f_{P\{1\}}$ and $\varphi_{P\{1\}}$ of Eqn. (S74) while all other parameters held constant for a $1P1B1C$ food web model. Note, the global optimum is located at a frequency of 1 d$^{-1}$; however, the peak is very narrow and was often missed by the hyperBOB search algorithm unless constraints on $f_{P\{1\}}$ were centered near 1.

For all simulations described in the main text, the following time varying square wave function for $\Omega_{1,P\{i\}}(t)$ was used instead of Eqn. (S74),

$$\Omega_{1,P\{i\}}(t) = \Omega_{amp\{i\}} \max\bigl(\delta_S\bigl(\mathrm{mod}(t,1), t_{On\{i\}}, \lambda_s\bigr) + \delta_S\bigl(\mathrm{mod}(t,1), t_{Off\{i\}}, -\lambda_s\bigr) - 1, 0\bigr), \quad \text{(S76)}$$

where $\delta_S(t, t_s, \lambda_s)$ is a smooth unit step function around $t_s$ given by,

$$\delta_S(t, t_s, \lambda_s) = \frac{1}{e^{-\lambda_s(t-t_s)} + 1}. \quad \text{(S77)}$$

In Eqn. (S76), three parameters, all bounded between 0 and 1, govern the characteristics of the $\Omega_{1,P\{i\}}(t)$ square-wave step function that occurs each day: $t_{On\{i\}}$ specifies time of day when the step-up occurs; $t_{Off\{i\}}$ specifies when the step-down occurs; $\Omega_{amp\{i\}}$ specifies the amplitude of the step. The parameter $\lambda_s$ was not considered a trait, but rather is used to control numerical smoothness of the step and was set to a value of 200 d$^{-1}$ for all simulations. For either function above, since biomass allocation must be conserved, $\Omega_{2,P\{i\}}(t)$ is obtained from the difference given by $1 - \Omega_{1,P\{i\}}(t)$.

One of the three types of temporal strategies discussed in the main text examines the impact of strict balanced growth for phytoplankton, so that the C:N ratio of phytoplankton remains constant (technically, this is the no-temporal-strategy strategy). To achieve balanced growth, $r_{1,P\{i\}}$ must be coupled to $r_{2,P\{i\}}$ so that the ratio of $S_{P\{i\}}$ concentration to $C_{P\{i\}}$ concentration remains constant, or that,

$$\frac{d\left[C_{P\{i\}}\right]/\left[\mathsf{S}_{P\{i\}}\right]}{dt} = 0, \tag{S78}$$

which occurs when,

$$\frac{r_{1,P\{i\}}(t)}{r_{2,P\{i\}}(t)} = \beta_{P\{i\}} \stackrel{\text{def}}{=} \left(\frac{1}{\varepsilon_{P\{i\}}} - n_{2,P\{i\}} + k_{PP\{i\}}\right), \tag{S79}$$

where $k_{PP\{i\}}$ is a specified constant that sets the ratio's $\left[C_{P\{i\}}\right]/\left[\mathsf{S}_{P\{i\}}\right]$ value. In simulations with balanced growth (i.e., no temporal strategy), $r_{1,P\{i\}}(t)$ and $r_{2,P\{i\}}(t)$ are calculated based on Eqns. (S12) and (S25), respectively, then adjusted as follows,

$$r_{1,P\{i\}}(t) = \min\left(r_{1,P\{i\}}(t), \beta_{P\{i\}} r_{2,P\{i\}}(t)\right) \tag{S80}$$

$$r_{2,P\{i\}}(t) = \min\left(r_{2,P\{i\}}(t), \frac{1}{\beta_{P\{i\}}} r_{1,P\{i\}}(t)\right). \tag{S81}$$

For example, at night, $r_{1,P\{i\}}(t)$ is zero and Eqn. (S81) forces $r_{2,P\{i\}}(t)$ to zero. Similarly, if no $NH_3$ is present, so that $r_{2,P\{i\}}(t)$ equals zero, then $r_{1,P\{i\}}(t)$ is set to zero based on Eqn. (S80) and phytoplankton dissipate solar radiation as particles, as described in Section S2.2.1 above.

Simulations in the main text were conducted with version 4.7 of the model, which can be obtained from GitHub (DOI: 10.5281/zenodo.3979922).

## S4 Example parameter input file

Below is the parameter file used for the circadian clock strategy at a dilution rate of 0.2 d$^{-1}$ using the nominal input concentrations given in Table 2 of the main text.

```
! Run149_opt4.5_1p1b1c
! Using AutoHetDet_Opt_V4.5
! 16-Jun-2020 on MEP
! This is Run149, but using V4.5.
! three parameters to specify a square wave function for omg_pp

¶ms
! Input parameters
npp  = 1 ! number of S1 primary producers
nbac = 1 ! number of bacteria
ncc  = 1 ! number of S2 consumers

! sumSigWeights determines which EP terms to use for optimization.
! Weights on EP where vector is: [Rxns, H20, particles]
! Total EP production use 1., 1., 1., for just rxn, use 1., 0., 0., etc.
sumSigWeights = 1., 1., 1.

! These parameters are used for EP surface generation only.
genSurf = .false. ! Should an EP surface be generated instead of optimization
```

```
readeps = .false. ! Read in the trait values from file.
whichPP = 1 ! which of the possible pp's to run f_pp and phi_pp over
nSurfPts = 1051  ! Number of points in the x and y dimension of the 2D surface to produce
reportTime = 10. ! How often to update screen during problem (min).

iseed = 10 ! changing this value to produce a different set of random values.
nuStar = 350.  ! Used in adaptive Monod equation 1/d
nuDet  = 175. ! For detritus decomp (1/d)
kappa = 5000. ! Adaptive Monod equation universal parameter (uM)
surA = 1.0 ! surface area of pond (m^2).
T_K = 293. ! get temperature (K)
pH  = 8.1  ! pH
depth = 1.0 ! Pond depth (m)
is = 0.72  ! Ionic strength (M) = 0.72*sal/35.0 (sal is salinity (PSU))
dil_t0 = 0.2 ! dilution rate at t0 (1/d)
dil_tf = 0.2 ! dilution rate at tf (1/d)
dil_n = 0    ! number of steps in dilution rate between t0 and tf

! Parameters associated with in-silico selection of traits
minCompFac = 500000.0 ! If process takes longer than (tf-t0)/minCompFac, then terminate
epp_min = 0.00001 ! min and max values for epp
epp_max = 1.0
ecc_min = 0.00001 ! min and max values for ecc
ecc_max = 1.0
ebac_min = 0.0001 ! min and max values for ebac
ebac_max = 1.0

! limits and parameters
! The square wave is limited to occur every day, so frequency is fixed in V4.0 and later
! tOn_pp is when the step up occurs, and tOff_pp when steps down occurs.  This are in days.
! Note, the overhangs (< 0 on tOn and >1 on tOff) insures omg_pp can be fully on all day
! because of the nature of the exp step function and value of sigOmg_pp.
! These are used for circadian strategy
sigOmg_pp = 200. ! this is used in the exp setup function to make a "smooth" square wave
tOn_pp_min  = -0.05 ! lower limit on on time (d)
tOn_pp_max  = 1.0   ! upper limit when step up can occur (d)
tOff_pp_min = 0.0   ! lower limit when a step down can occur (d)
tOff_pp_max = 1.05  ! upper limit when step down occurs (d)
! Use these for no circadian rhythm (i.e., passive storage).
! tOn_pp_min  = -0.05 ! lower limit on on time (d)
! tOn_pp_max  = -0.049   ! upper limit when step up can occur (d)
! tOff_pp_min = 1.04   ! lower limit when a step down can occur (d)
! tOff_pp_max = 1.05  ! upper limit when step down occurs (d)

! V4.7 Add binaryOMG to set omg_cc to binary matrix (only 0's or 1's)
binaryOMG = .false. ! default is .false.

! Coupling between r_1,p and r_2,p.  If k_pp below is set to zero (default) then
! reactions are not coupled, but if k_pp > 0, then it sets the ratio of
! C_p to p (i.e., C_p/p = k_pp). Note, there is also the variable k_p for light
! attenuation that is different. When k_pp /= 0, tOn_pp and tOff_pp should be set to
! the passive storage scenario.
k_pp = 0.0
```

```
! Initial and feed concentrations in input feed. All concentrations in uM
dicI = 2000. ! (uM)
o2I = 225. ! (uM)
nh3I_t0 = 5. ! nh3I at t0 (uM)
nh3I_tf = 5. ! nh3I at tf (uM)
nh3I_n  = 0   ! Number of steps in nh3I between t0 and tf
c_LI = 10. ! labile carbon (uM)
c_dI = 100.0 ! detrital carbon (uM)
n_dI = 7. ! detrital nitrogen (uM)
ppI  = 0.1 ! initialize all phytoplankton to this value (uM)
c_ppI = 0.1 ! all phytoplankton carbon stores (uM)
ccI   = 0.1 ! initialize all consumers to this value (uM)
bacI = 0.1 ! all bacteria (uM)

! Phosphate concentrations. Held fixed, but used for thermodynamic calculations
h3po4 = 1. ! uM
P_d   = 5. ! detrital P (uM)

! Biomass elemental composition. From Battley1998 for yeast. Unit carbon
alf = 1.613 ! H
bet = 0.557 ! O
gam = 0.158 ! N
del = 0.012 ! P
cell_F = 1000.0 ! concentration factor for C_P. Intracellular versus extracellular volume.
delPsi = 0.1 ! LaRowe's thermo driver.  Cell membrane potential (V)

! Gas Exchange, temp and pH
pV_o2  = 3.0 ! piston velocity for O2 (m/d)
pV_co2 = 2.6 ! piston velocity for CO2 (m/d)
pO2    = 0.21    ! partial pressure of O2 in atmosphere (atm)
pCO2   = 400.d-6 ! partial pressures CO2 in the atmosphere (atm)

! Solar parameters
I0max = 406000. ! Solar constant in PAR (mmol photons /m^2 /d)
dLat = 42.0 ! Latitude for calculating solar radiation
dGr_Ggamma = -253. ! Gibbs free energy of photons (J/mmol photon of blue light, 440 nm)
k_w = 0.011   ! water attenuation coef (1/m)   (see Table 2.11 of Wozniak2007 for 430 nm light)
k_p = 0.000625 ! attenuation coef by non-algal parties   (m^2/mmol-S)
k_chla = 0.0025 ! attenuation coef by algal pigments. (Wozniak2007, adjusted to (m^2/mmol-S))

! BiM ODE solution parameters
ompThreads = 1 ! Specify how many threads to use (no currently used, set to 1)
t0 = 0. ! Start time for ODE integration (d)
tDays = 730. ! number of days to run simulation (d)
t0_ep = 0.   ! For optimization, interval over which EP production is maximized.
tf_ep = 730.  ! end of EP interval.
maxstep_BiM = 10000000 ! maximum number of BiM iterations (set to 0 to use default of 100000)
useOmpJac = 0 ! set to 0 to have BiM calculate numerical gradient
maxattempts = 1 ! number of attempts to solve ODEs before declaring failure
absZero = 1.e-8 ! Numbers less than this are set to this value smoothly. (div by 0 prevention)
atol1 = 1.0e-6  ! absolute tolerance for BiM
rtol1 = 1.0e-6  ! relative tolerance for BiM
hmax_BiM = 0.05 ! largest step size (d). default = (TEND-T0)/8
```

```
! parameters used by hyperBOB
rhobeg = 0.49 ! initial and final values of a trust region radius
rhoend = 0.0001 ! When trust region is less than rhoend, stop.
iprint = 0 ! controls amount of printing (0, 1, 2 or 3)
maxfun = 10000 ! maximum number of calls to CALFUN
optimize = .true. ! If true, MEP optimization occurs, otherwise just solve ODEs
fcnUpdate = 100 ! output current status after every fcnUpdate ODE integrations
/
```